\documentclass[iop]{emulateapj-rtx4} % read by: ACROREAD
\shortauthors{Sekanina}
\shorttitle{Comet Pair C/1844 Y1 and C/2019 Y4}
\slugcomment{Version \today }

\begin{document}
\title{Fragmentation and History of Comet Pair C/1844 Y1 and C/2019 Y4\\[-1.6cm]}
\author{Zdenek Sekanina}
\affil{La Canada Flintridge, California 91011, U.S.A.; \it ZdenSek@gmail.com}
% \email{}
\begin{abstract} % maximum length = 1920 characters
I call attention to extraordinary features displayed by the genetically
related long-period comet pair of C/1844~Y1 (Great Comet) and C/2019~Y4
(ATLAS).  The issue addressed most extensively is the fragmentation and
disintegration of the latter object, itself a thousands-of-years-old
fragment.  Of the four fragments of C/2019~Y4 recognized by the Minor
Planet Center --- A, B, C, and D --- I confirm that B was the
principal mass, which stayed undetected until early April.  The
comet's 2020 fragmentation is proposed to have begun with a separation
of B and A near 22~January, when the nuclear condensation suddenly
started to brighten rapidly.  From late January to early April, only
Fragment A was observed.  The remaining Fragments C and D split off
most probably from A in mid-March, but they too were detected only in
April.~A new fragment, E, is proposed to have~been~observed on three
days.  Further addressed are the issues of the orbital period and antitail
of C/1844~Y1 and the extreme position of this pair among the genetically
related groups/pairs of long-period comets.
\end{abstract}
\keywords{individual comets: C/1844 Y1, C/2019 Y4; methods: data analysis}
\section{Introduction} %%% Sec. 1
%
% \vspace{-0.08cm}
%
The arrival of comet C/2019 Y4 a few years ago~was~the most recent addition
to the category of pairs or small groupings of common-origin long-period
comets, with the members passing perihelion months, years, decades, or
even centuries apart.  Discovered by Project ATLAS, {\it Asteroid
Terrestrial-Impact Last Alert System\/}, the 2019 comet was a distant
companion to the brilliant comet C/1844~Y1 (Green 2020a), having
apparently~broken all records by returning to perihelion not
until~175.5~yr~later.  As fragments of a common parent, they are believed~to
have been its parts until~about~the~time~of~\mbox{perihelion} passage several
thousand~years ago.  The time span~of almost two centuries between the two
fragments'~arriv\-als dwarfs those of only 8 to 32~years exhibited by~the
three companions to C/1988~A1 (Liller), the leading bright
member of the best known comet group of this kind.  The recent
fragment of that group, C/2019~Y1~(\mbox{ATLAS}), was by sheer coincidence
discovered by the same project (Green 2020b) only twelve days before
C/2019~Y4.  In Section~7 I will return to the relationships among the
long-period comets' companions in general and between C/1844~Y1 and
C/2019~Y4 in particular.

The early recognition by M.\ Meyer of the near identity between the
orbits of C/2019~Y4 and C/1844~Y1 (Green 2020a) was enough to turn the
former object~into~a~likely candidate for preperihelion disintegration.
It is known that most companions of the split comets have fairly short
lifetimes, which statistically vary as an inverse 0.4 power of the
nongravitational acceleration (Sekanina 1982).  Even though scatter is
rather large, for{\vspace{-0.085cm}} nearly-parabolic comets the lifetime
tends to vary as $q^{-\frac{1}{2}}$, and it can be demonstrated that for
\mbox{$q = 0.25$ AU} only massive fragments (with a nongravitational
{\vspace{-0.04cm}}acceleration lower than 3~units of 10$^{-5}$\,the
solar gravitational acceleration, or approximately 10$^{-8}$\,AU
day$^{-2}$ at heliocentric distance of 1~AU) should survive one full
revolution about the Sun.  This condition was apparently satisfied by
C/1844~Y1, but not by C/2019~Y4.

\section{The Light Curve of C/2019 Y4}
The monitoring of brightness variations, as the often readily available
signature of a comet, is very useful in order to outline and understand
its activity even when the object is evolving in a normal, fairly
foreseeable fashion; it is absolutely essential when the object is
subjected to irregular, explosive processes.  For any comet under
observation, information on the brightness evolution of the nuclear
condensation is customarily provided online by observers alongside
the CCD astrometric data in the {\it Minor Planet Electronic
Circulars\/} (MPEC) and later on in the {\it Minor Planet Circulars\/},
both issued by the {\it Minor Planet Center\/} (MPC).  The total
visual or CCD magnitudes are conveniently available online from
the {\it International Comet Quarterly\/} (ICQ) or the Crni Vrh
Observatory's {\it Comet Observation Database\/} (COBS).

Comet C/2019 Y4 had been discovered on 28 December 2019, but subsequently
a pair of its pre-discovery images was spotted in a {\it Catalina Sky
Survey\/} field taken on 20~December (e.g., MPC Staff 2020).  A set of
selected apparent CCD magnitudes of the nucleus condensation reported
between 20~December and 25~February 2020 is plotted in Figure~1.  Each
data point is an average of such magnitudes obtained on a given night at
a given observing site.  Practically the same curve, with only moderately
increased scatter, would result, if all available data were plotted
instead.

\begin{figure}[t]
\vspace{0.2cm}
\hspace{-0.18cm}
\centerline{
\scalebox{0.68}{                   
\includegraphics{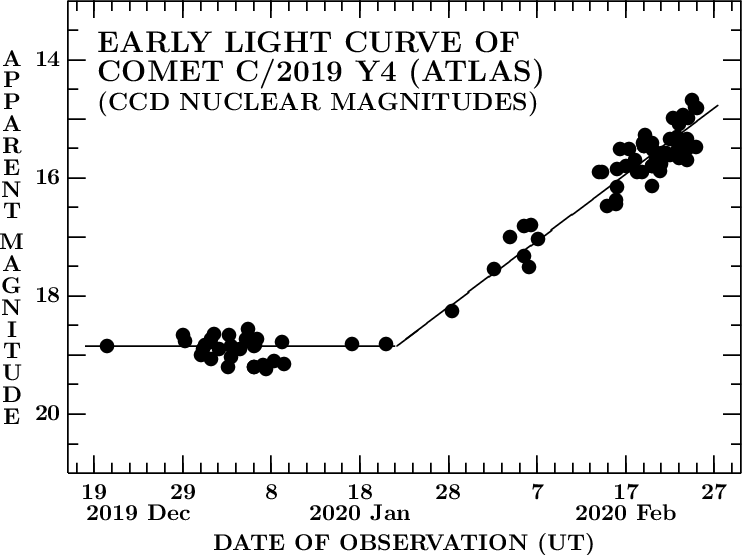}}} % from f1_atlas.tex   FIGURE 1
\vspace{-0.1cm}
\caption{Early light curve of the nuclear condensation of Comet C/2019~Y4
 based on a set of selected CCD observations between 20~December 2019 and
 25~February 2020.  The curve shows no trend until about 22~January, at
 which time the brightness began to increase exponentially (linearly on
 the magnitude scale).  The data are from MPEC 2020-K131.{\vspace{0.6cm}}}
\end{figure}

The remarkable feature of the figure is a sudden change in the slope of
the light curve around 22~January.  The apparent nuclear brightness,
essentially constant before this date, began to rise at an essentially
constant rate of about 0.11~mag per day afterwards.  This behavior of
the light curve is entirely unlike a typical outburst, because the
rate of brightening is much too slow and the duration of brightening
much too long; instead of a fraction of a day or a few days at most,
the brightness was continuing to rise over a period of longer than a
month, with no sign of ebbing. I will return to the significance of
this phenomenon in Section 3.{\pagebreak}

By the end of February the comet became bright enough that visual
observers could begin to apply their methods of total-brightness
determination.  Examples of light curves, based on such fairly
systematic data sets, as well as on large-aperture CCD imaging,
are displayed in Figure~2 from 25~February 2020 on.  These light
curves were generated independently of the efforts directed towards
the comet's astrometry.

Cursory inspection of the four curves shows both commonalities and
differences.  They all indicate the comet to have been brightening up
to \mbox{25--29}~March, the peak rates reaching 0.20\,$\pm$\,0.07~mag
per day.  The CCD curves show the peak brightness fainter than
magnitude 8, whereas the visual curves indicate it was between 7 and
8; this only may mean that the visual observers were accounting for
the outer regions of the coma, which were outside the CCD imaging
apertures.  The brightness decline during April appears to have been
very modest and the minimum even broader than the late-March peak.
The most complete light curve in Figure~2, by T.~Lehmann, shows that
in May the comet slowly brightened again, while C.~Harder's results
suggest the opposite.  The comet was last seen on 21~May, 10~days
before perihelion.

\begin{figure}[t]
\vspace{0.19cm}
\hspace{-0.2cm}
\centerline{
\scalebox{0.615}{
\includegraphics{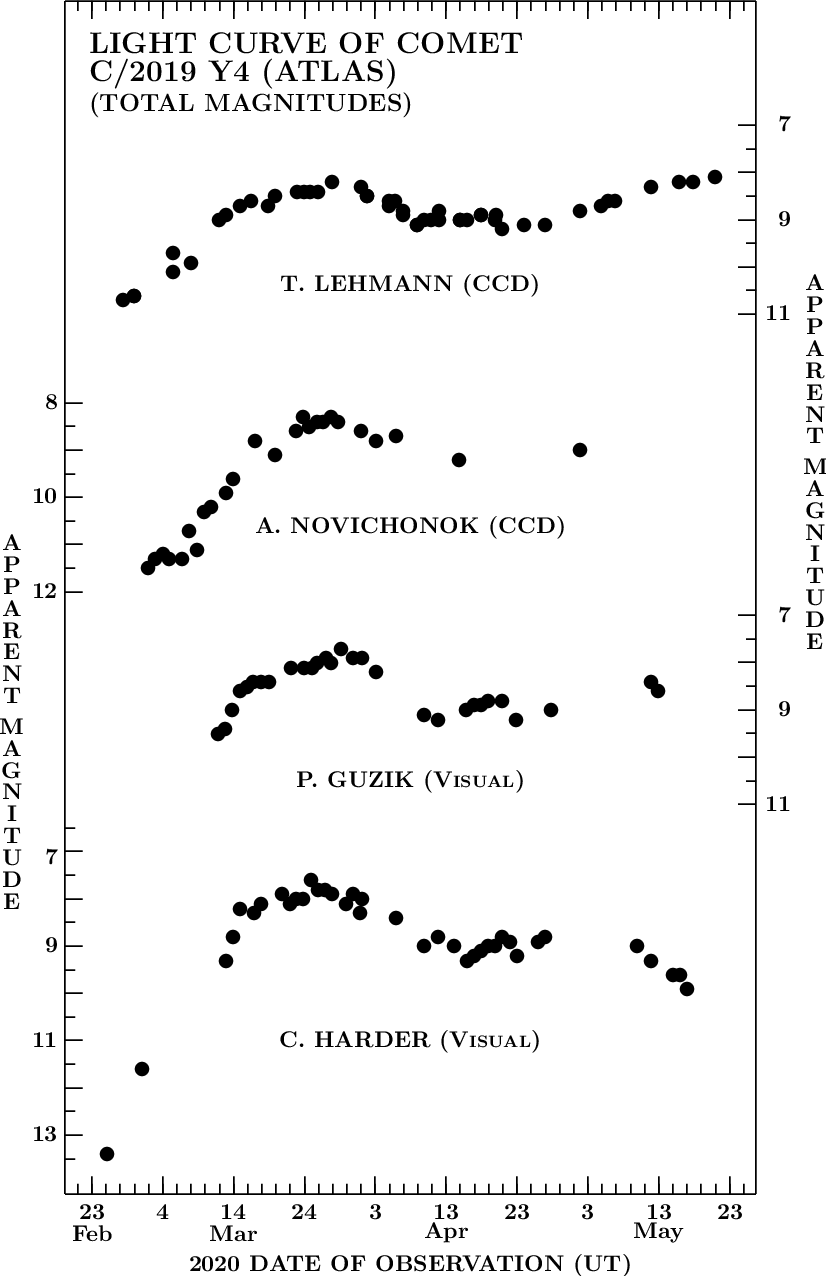}}} %   FIGURE 2
\vspace{-0.1cm}
\caption{Light curves of Comet C/2019 Y4 (total magnitudes) by
 two CCD and two visual observers; around 25~February the total
 brightness is estimated at about \mbox{2--3}~magnitudes higher
 than the nuclear brightness in Figure~1.  The rate of brightening
 from that time on, up to late March, is even steeper than the
 rate exhibited by the data for the nuclear condensation before
 25 February.{\vspace{0.5cm}}}
\end{figure}

\section{Nuclear Fragmentation of C/2019 Y4}
The earliest indirect evidence of possible splitting of the nucleus was
published on 5~April, when S.~Nakano reported detection of the sudden
appearance of nongravitational perturbations of the comet's orbital
motion (Green 2020c).  A day later Ye \& Zhang (2020) suggested that
the comet may be disintegrating, as images taken on 5~April with a
60-cm telescope of the Xingming Observatory showed an ``elongated
pseudo-nucleus.'' On 7~April a secondary feature was found by E.\ Guido
et al.\ to be embedded in the tail several arcseconds from the nuclear
condensation and then two more features in images taken at the Celado
Observatory (Green 2020d).  By 13~April CCD images obtained by many
observers showed that this comet had unquestionably fragmented.

To examine the time sequence of the nuclear fragmentation of this comet,
I adopt the classification of fragments by the MPC Staff (2020),
according to which up to four condensations --- A, B, C, and D --- could
be measured from the Earth; Fragment~B was proposed as the principal mass.
This identity is being confirmed by my dynamical modeling below.  The MPC
Staff similarly established that the object which had provisionally been
designated as Fragment~E, was subsequently identified as Fragment~B.

\begin{table}[t]
\vspace{0.13cm}
\hspace{-0.19cm}
\centerline{
\scalebox{0.96}{ 
\includegraphics{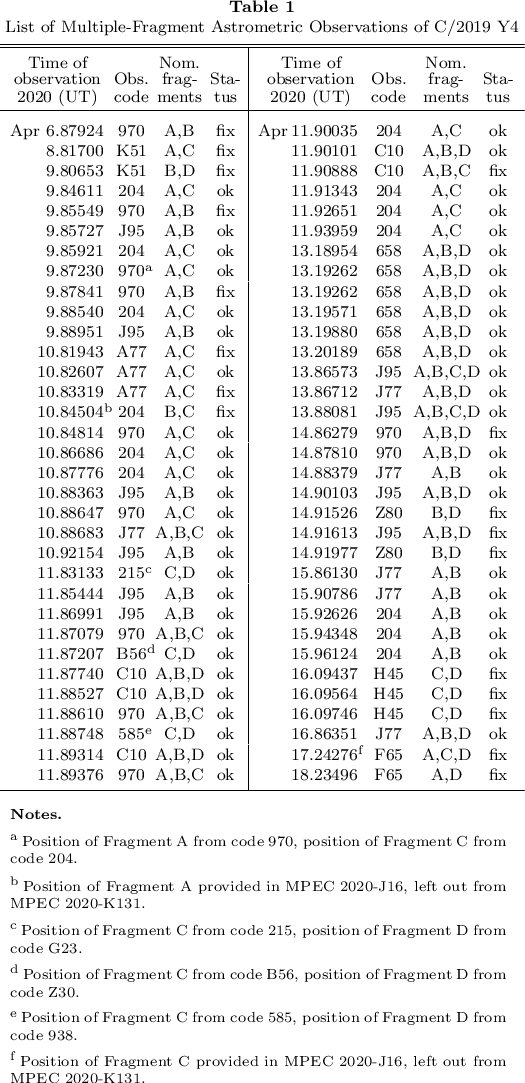}}} %  TABLE 1
\vspace{0.6cm}
\end{table}

Application of the fragmentation model that I introduced nearly 50~years
ago (Sekanina 1977) and its expanded version (Sekanina 1978) allow one
to obtain multiparametric solutions from sets of separation vectors of any
two fragments.  Since the astrometric data published for comet C/2019~Y4
are the fragments' equatorial coordinates (MPC Staff 2020), the relevant
input for the proposed modeling is provided by the astrometric positions
of fragment pairs at the given times.  As the first step, the instances
and fragments for which such data are available are summarized in Table~1.
The columns list the time of observation, the standard observatory code,
the fragments assigned to the measured astrometric positions by the MPC
Staff, and the status, describing the outcome of preliminary tests aimed
at finding out which fragments satisfy performed solutions and which do
not.  The {\it ok\/} status acknowledges that the assigned fragments
offer consistent results, while the {\it fix\/} status warns that
revisions are necessary.

\begin{table}[t]
\vspace{0.15cm}
\hspace{-0.22cm}
\centerline{
\scalebox{0.965}{ 
\includegraphics{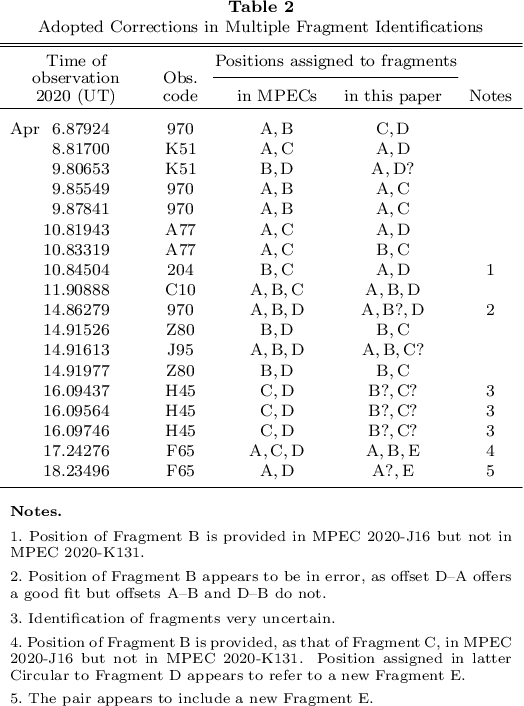}}} % t0_atlas.tex  TABLE 2
\vspace{0.6cm}
\end{table}

The required revisions, presented in Table~2, are reassigned fragments
to the astrometric positions that did not fit the preliminary
fragmentation solutions.  Such revisions are rather straightforward
when only two fragments are involved.  In cases of more than two
fragments the situation could get more confusing.  For example, in
a configuration of three fragments A, B, and C or D, the third one
may better fit as C relative to A, but as D relative to B.  Since
a fragment cannot be both C and D, a difficult decision has to be
made, based for example on the magnitude of the residuals.

In the following I address the relationships among the four fragments.
From Tables~1 and 2 it follows that to the terrestrial observers the
most extended period of time during which the multiple nucleus can be
investigated using the proposed method is \mbox{6--18}~April, less
than two weeks.  During this period, the separation distance between
Fragments~A and B varied rather insignificantly, while Fragments~C and
especially D were gradually receding from both A and B.  Under the
circumstances, the most feasible strategy appears to be to investigate
first the relation between A and B, and only later the relation of C
and D with respect to A and B.  Both C and D rapidly grew more diffuse
and were obviously disintegrating into expanding clumps of scattered
material before the eyes of the observers.

\begin{table*}[t]
\vspace{0.2cm}
\hspace{-0.2cm}
\centerline{
\scalebox{1.02}{
\includegraphics{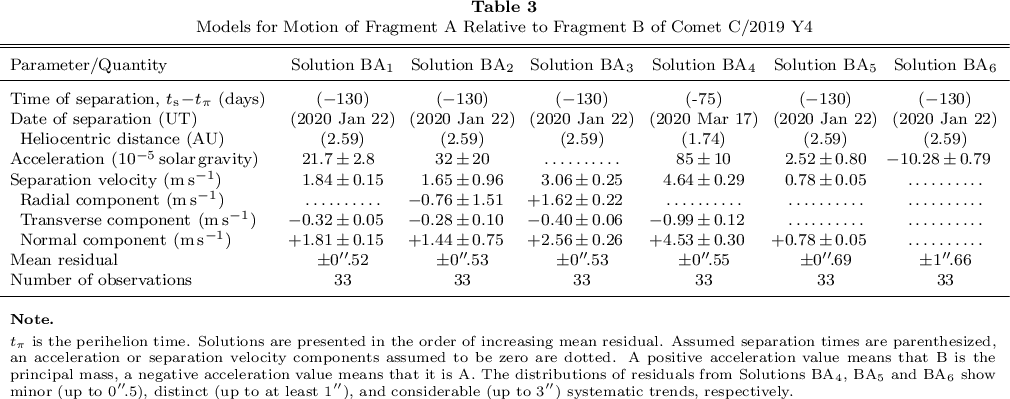}}} %  TABLE 3
\vspace{0.53cm}
\end{table*}

Imaged by the Hubble Space Telescope (HST) on 20 and 23 April,
Fragments~A and B appeared as clusters of debris as well (Ye et al.\
2021).  I will not address the HST appearance of the comet in this
paper beyond remarking that the considerable degree of fragmentation
of the nucleus seen was likely to be due not only to the HST's superior
spatial resolution and detection limit, but also because of the late
stage of the rapidly advancing disintegration process.

The model applied in the following is described in detail in Sekanina
(1977, 1978) and its features are summarized in Sekanina (1982).
Here I only note that the model allows one to determine up to five
parameters:\ the time of separation, the nongravitational
acceleration (actually the companion, or secondary mass, is always
{\it de\/}celerated relative to the principal mass), and, if needed,
the three components (radial, transverse, and normal) of the separation
velocity of the companion from the principal~mass.  The acceleration's
variation with heliocentric distance is approximated by an inverse square
power law and the magnitude is usually expressed in units of 10$^{-5}$\,the
solar gravitational acceleration.  In practice, the parameters are
determined by a least-squares differential-correction optimization
procedure, but the user is allowed to attach any value(s) to any one or
more of the five parameters and circumvent its optimization if he chooses
to do so (or has to in order to maintain the convergence).  This is an
extremely useful feature, which implies~that the model is effectively
provided in 31~different versions.

The optimized differential nongravitational acceleration as the model's
primary driver for separating fragments is linked to another important
feature:\ its sign automatically determines which of the two tested
fragments, X or Y, is the principal one.  It makes no difference whether
one tests a set of separation vectors \mbox{Y--X} or \mbox{X--Y}.  If
a model solution derived from the \mbox{Y--X} set yields a positive
acceleration, X is the principal fragment and vice versa.  Similarly
with the set \mbox{X--Y}.

\subsection{Fragments A and B}
Inspection of the employed data, reported by the MPC and summarized in
Table~1, indicates that Fragments~A and B were simultaneously imaged
38~times in the period of \mbox{6--18 April}, but the results of
preliminary tests presented in Table~2 suggest that on a few occasions
at least one of the two fragments was either misidentified or its
position measured too inaccurately to be of any use in the model
computations.  Adopting a rejection cutoff of $\pm$1$^{\prime\prime}\!$.5
for the residuals $O\!-\!C$ (observed minus computed from the model)
of \mbox{A--B} in right ascension and declination, the number of
acceptable observations has turned out to equal 33.

My efforts were first directed toward the simplest, two-parameter solution,
trying to determine the time of separation of Fragment~A from Fragment~B
and the nongravitational acceleration of A relative to B.  I began with
a separation time of 75~days before perihelion, based on Hui \& Ye's
(2020) conclusion that ``the comet has disintegrated since 2020 mid-March.''
Unfortunately, after a few initial iterations suggesting a meaningless
separation time as far back in time as some 600~days (sic) before
perihelion, the differential-correction optimization procedure failed
to converge.  Similarly disappointing were the successive attempts at
solutions with more than two parameters.

The overall impression from these runs was that even though the separation
time appeared to be indeterminate, the solutions in which the separation
was {\it assumed\/} to have taken place much earlier than $\sim$100~days
before perihelion (say, prior to February 2020) looked distinctly preferable,
offering a lower mean residual and better distribution of individual
residuals.  

Given that splitting of a comet's nucleus is known to often coincide with
suddenly elevated activity showing up as a brightening in the light curve,
the abrupt change in the slope around 22~January, or 130~days before
perihelion, in Figure~1 is a strong suspect as a potential candidate for
the fragmentation event that gave birth to components A and B.  Incorporating
this assumption, the one-parameter solution fitted the observations with
an unacceptably high mean residual and with the individual residuals
displaying strong systematic trends of up to some 3$^{\prime\prime}$.
Solving in addition for the normal component of the separation velocity
improved the solution's quality (BA$_5$ in Table~3), the systematic trends
in the distribution of residuals now reaching not more than about
1$^{\prime\prime}\!$.

\begin{table}[t]
\vspace{0.11cm}
\hspace{-0.17cm}
\centerline{
\scalebox{0.965}{
\includegraphics{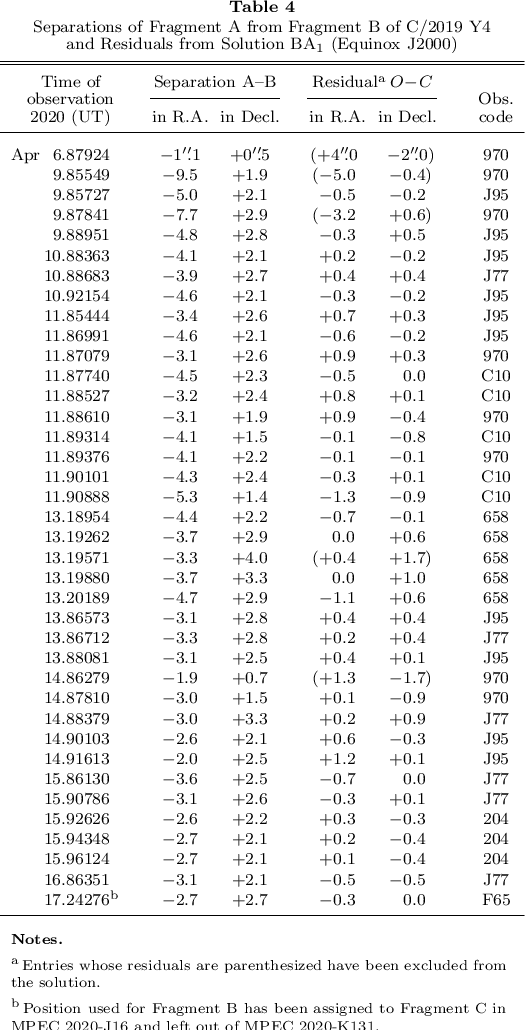}}} %  TABLE 4
\vspace{0.7cm}
\end{table}

The most satisfactory fit to the data was offered by a three-parameter
solution, presented in Table~3 as BA$_1$.  The acceleration slightly exceeds
20 units of 10$^{-5}$\,the solar gravitational acceleration and the
separation velocity stays below 2~m~s$^{-1}$.  The table shows that
this solution is distinctly preferable to BA$_4$, an equivalent solution
for an {\it assumed\/} separation time of 75~days before perihelion
(mid-March), which{\vspace{-0.03cm}} requires a high separation velocity
of nearly 5~m~s$^{-1}$ and an acceleration of companion~A that is four times
higher.  Solution BA$_1$ is also preferable to BA$_3$, in which the
contribution from the acceleration is taken over by the (increased)
separation velocity.  On the other hand, a four-parameter solution
BA$_2$ that incorporates the radial component of the separation velocity
among the parameters is a case of overkill, documented by their unacceptably
high errors.

\begin{figure}[t]
\vspace{0.14cm}
\hspace{-0.2cm}
\centerline{
\scalebox{0.59}{
\includegraphics{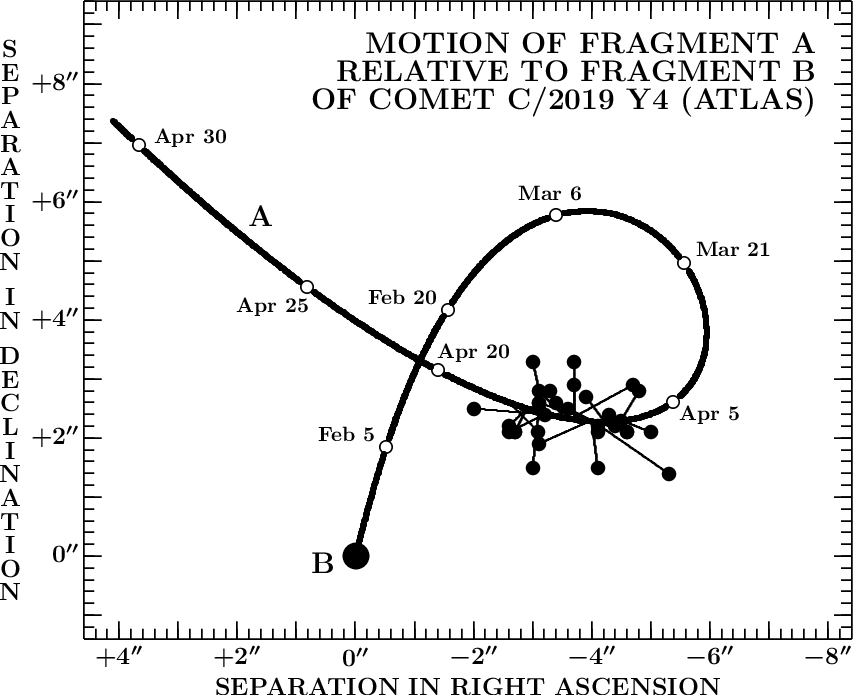}}} %  FIGURE 3
\vspace{-0.03cm}
\caption{Separation trajectory of Fragment A relative to Fragment~B
derived from Solution BA$_1$ in Table~3.  The solid circles are the
observations of A from Table~4, while the open circles are ephemeris
positions (at 0~TT) from the model.  The two fragments are assumed to
have separated from each other on 22~January 2020, 130~days before
perihelion.  Note that the nucleus was not seen double during March,
even though the separation distance between A and B was greater than
in April.{\vspace{0.7cm}}}
\end{figure}

The residuals from the fit of Solution BA$_1$ of Table~3 to the
38~\mbox{A--B} separations in right ascension and declination
are listed in Table~4; of the five rejected data points three are
incorrect identities (Table~2) and two are inaccurate data.  The
peculiar separation trajectory of Fragment~A relative to B, plotted
in Figure~3, is consistent with a scenario that incorporates the
fragmentation event in late January and is, in addition, supported
by the nuclear-condensation's light curve in Figure~1 and by the fact
that the nucleus was not seen double during February and especially
in March 2020, when the predicted separation distance between the two
fragments was much greater than in April (Figure~3).

The key to understanding both the light curve and the detection anomaly
is the fundamentally different behavior of the two fragments:\ while
Fragment~A was subjected to a high rate of mass loss, {\it Fragment~B
was\/} of low activity and {\it too faint to detect\/} in any telescope
employed in the comet's observations.  The observed light curve between
late January and early April was exclusively a product of Fragment~A.

Closer inspection of the light curve shows that the constant apparent
brightness of the nuclear condensation between late December 2019 and
late January 2020 actually meant a decline of activity.  If interpreted
in terms of the cross-sectional area of dust grains near the nucleus,
the drop was from 96~km$^2$ on 20~December to 33~km$^2$ on 22~January
at an assumed geometric albedo of 4~percent and accounting for the
phase effect using the Marcus (2007) law.  The observed brightness
increase from apparent magnitude of 18.8 on 22~January to 15.0 on
25~February, at a fairly constant rate of 0.11~mag per day is equivalent
to an {\it exponentially\/} growing cross-sectional area of released
dust between the two dates from 33~km$^2$ to 367~km$^2$.  At time $t$
this cross-sectional area equals
\begin{equation}
{\cal A}(t) = {\cal A}(t_{\rm frg}) \exp \left[ \zeta (t \!-\! t_{\rm frg})
 \right],
\end{equation}
where $t_{\rm frg}$ is the presumed fragmentation time, 22~January.  As
\mbox{$\exp ( 34 \,\zeta ) = 367/33 = 11.1$}, one finds \mbox{$\zeta =
0.071$} per day.  The comet brightening continuously at the~observed rate
of 0.11~mag per day and its description~via Equation~(1) is thus readily
interpreted as progressive disintegration of Fragment~A, whose cross-sectional
area doubled in approximately 10~days.

\begin{table*}
\vspace{0.15cm}
\hspace{-0.23cm}
\centerline{
\scalebox{1}{
\includegraphics{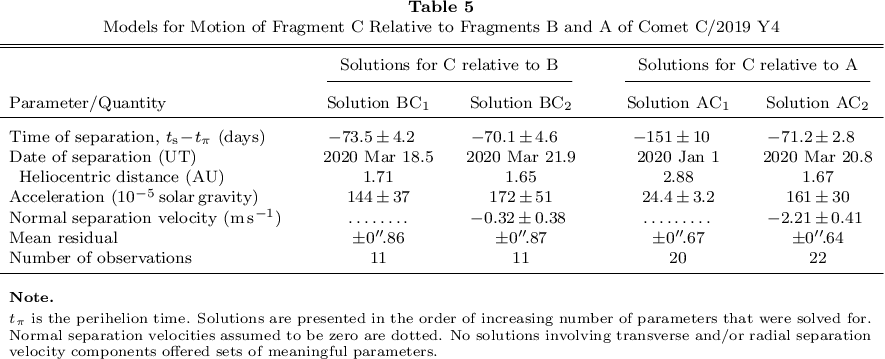}}} %  TABLE 5
\vspace{0.5cm}
\end{table*}

To properly investigate the crumbling of Fragment~A that presumably
began on 22~January, one should disregard the presence of dust in
the atmosphere at that time and ${\cal A}_{\rm frg}$ in Equation~(1)
should be the cross-sectional area of the surface of Fragment~A.  The value
of $\zeta$ would then be somewhat higher.  For example, taking conservatively
the diameter of Fragment~A at the time of separation from Fragment~B
to be 500~meters, one would obtain \mbox{$\exp(34\,\zeta) = 4
\times$367/($\pi \!\times\! 0.5^2) \simeq 1870$} or \mbox{$\zeta =
0.22$} and the doubling time a little over 3~days.  For a fragment
100~meters across, \mbox{$\zeta = 0.32$} and the doubling time
more than 2~days.

Continuing with this line of argumentation,
let at time \mbox{$t > t_{\rm frg}$} the number of pieces of debris of
the crumbling fragment be ${\cal N}(t)$ and the mean cross-sectional
area of one piece $\langle {\cal A} \rangle$, so that
\begin{equation}
\langle {\cal A} \rangle \: {\cal N}\:\!\!(t) = {\cal A}(t) = 
 {\cal A}(t_{\rm frg}) \,\eta(t),
\end{equation}
where \mbox{$\eta(t) = \exp[\zeta (t \!-\! t_{\rm frg})]$}.  Expressing a
cross-sectional area in terms of a characteristic dimension $\Re(t)$,
such as the radius or diameter, one can write the conditions for the
mean cross-sectional area and volume at time $t$:
\begin{eqnarray}
\langle \Re^2 \rangle \, {\cal N} & = & \Re_{\rm frg}^2 \, \eta(t),
 \nonumber \\[-0.05cm]
\langle \Re^3 \rangle \, {\cal N} & = & \Re_{\rm frg}^3,
\end{eqnarray}
where $\Re_{\rm frg}^2$ and $\Re_{\rm frg}^3$ refer to Fragment~A
at the fragmentation time.  The ratio,
\begin{equation}
\frac{\langle \Re^3 \rangle}{\langle \Re^2 \rangle} = \frac{\Re_{\rm
 frg}}{\eta(t)},
\end{equation}
is a function of the size distribution, \mbox{$f(\Re) \, d \Re \sim
\Re^{-s} d \Re$}, of the debris.  If the minimum and maximum dimensions
at time $t$ are $\Re_{\rm min}(t)$ and $\Re_{\rm max}(t)$, respectively,
the mean values are given by
\begin{equation}
\langle \Re^\nu \rangle \!\! \int_{\Re_{\rm min}}^{\Re_{\rm max}} \!\!
 \Re^{-s} d \Re = \!\! \int_{\Re_{\rm min}}^{\Re_{\rm max}} \!\!
 \Re^{\nu-s} d \Re \;\;\;\: (\nu = 2, 3).
\end{equation}
For less steep distributions, for which \mbox{$s < 3$} the mean values
are essentially independent of $\Re_{\rm min}$ and the ratio is at time
$t$
\begin{equation}
\frac{\langle \Re^3 \rangle}{\langle \Re^2 \rangle} = \frac{3\!-\!s}{4\!-\!s}
 \, \Re_{\rm max}(t). \\[0.01cm]
\end{equation}
Comparison of Equations (4) and (6) gives for the dimension of the largest
piece of debris of Fragment~A at time $t$ in terms of the fragment's
dimension at $t_{\rm frg}$:
\begin{equation}
\Re_{\rm max}(t) = \frac{4 \!-\! s}{3 \!-\! s} \,
 \frac{\Re_{\rm frg}}{\eta(t)} = \frac{4 \!-\!s}{3 \!-\! s} \,
 \Re_{\rm frg} \exp[-\zeta (t \!-\! t_{\rm frg})].
\end{equation}
Regardless of exact values for $\zeta$ and $s$ (if lower than 3),
Equation~(7) suggests that months after the separation of A from B, the
dimensions of the largest remaining piece of A should be orders of
magnitude smaller than the initial size.  This is contrary to the
results that Ye et al.\ (2021) obtained from their study of the
comet's debris with the HST in late April.  It appears that the
process of disintegration of Fragment~A was proceeding in a highly
irregular manner.  The rate of disintegration may have varied strongly
with time and/or the process affected only a fraction of the object.
In any case, the presence of major anomalies is implied by the dramatic
fluctuations in the size distribution of the debris of Fragment~A, as
documented by Ye et al.

\subsection{Fragment C and Its Parent}
The statistics of the split comets with more than two fragments suggest
that, as a rule, the secondary nuclei (or companions) have separated from
the principal mass (Sekanina 1982).  Less often has a second companion
split off instead from the first companion, or, more generally, a
companion of a later generation from a companion of an earlier generation.
Accordingly, one would expect Fragment~C of comet ATLAS to separate,
like Fragment~A, from B.  Yet, it is desirable to test both options.
An obvious constraint is that Fragment~C could {\it not\/} have broken
off from A {\it before\/} A did from B.

\begin{table}[t]
\vspace{0.1cm}
\hspace{-0.2cm}
\centerline{
\scalebox{0.97}{
\includegraphics{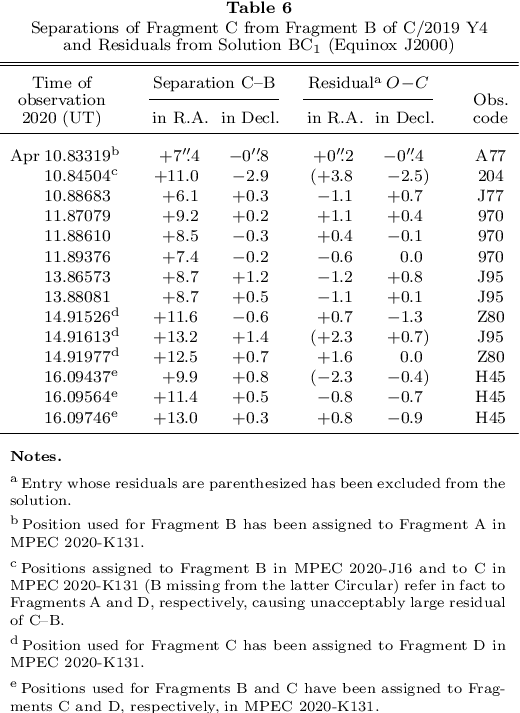}}} %  TABLE 6
\vspace{0.5cm}
\end{table}

Tables 1 and 2 show that the simultaneous astrometric observations of B
and C have been less numerous than those of A and C, and both pairs less
numerous than the simultaneous astrometry of B and A.  Because of these
limitations, the rejection cutoff for examining the parent of C was
increased from 1$^{\prime\prime}\!$.5 to 2$^{\prime\prime}\!$.0.  I
applied a variety of the fragmentation model's versions and found that
unlike in the case of the motion of Fragment~A relative to B, the
separation time always converged.  However, solutions with higher
number of parameters than three consistently failed to converge; the
two converging ones for either fragmentation scenario are compared in
Table~5.

Solutions~BC$_1$ and BC$_2$ offer similar separation times, which
precede the peak of the light curves in Figure~2 by several days,
but Solution~BC$_1$ is preferred because the normal component of the
separation velocity from Solution~BC$_2$ is poorly determined and
therefore an unnecessary parameter.  An unexpected problem is a
systematic trend in the residuals in declination, which mars {\it
either\/} solution.  For Solution~BC$_1$ this undesirable anomaly
is apparent from Table~6 and very obvious from the separation
trajectory of Fragment~C relative to Fragment~B displayed in
Figure~4.

\begin{figure}[b]
\vspace{0.8cm}
\hspace{-0.22cm}
\centerline{
\scalebox{0.633}{
\includegraphics{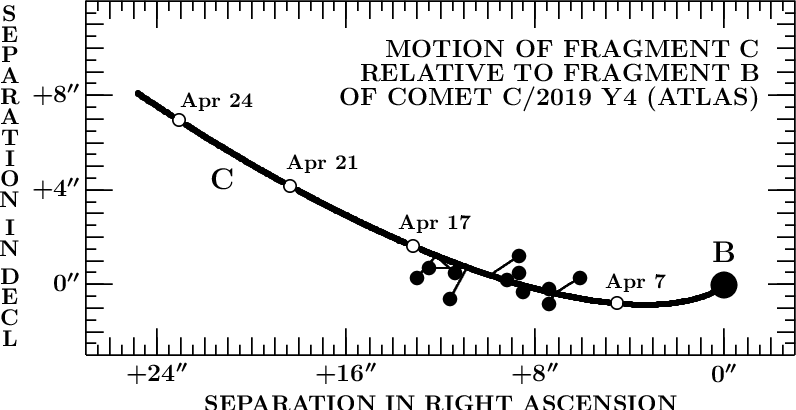}}} %  FIGURE 4
\vspace{-0.05cm}
\caption{Separation trajectory of Fragment C relative to Fragment B derived
from Solution BC$_1$ in Table 5.  The solid circles are the observations of
C from Table 6, while the open circles are ephemeris positions (at 0 TT)
from the model.  Note the systematic trend in the distribution of the data
relative to the model curve.{\vspace{-0.04cm}}}
\end{figure}
\begin{table}[t]
\vspace{0.1cm}
\hspace{-0.23cm}
\centerline{
\scalebox{0.97}{
\includegraphics{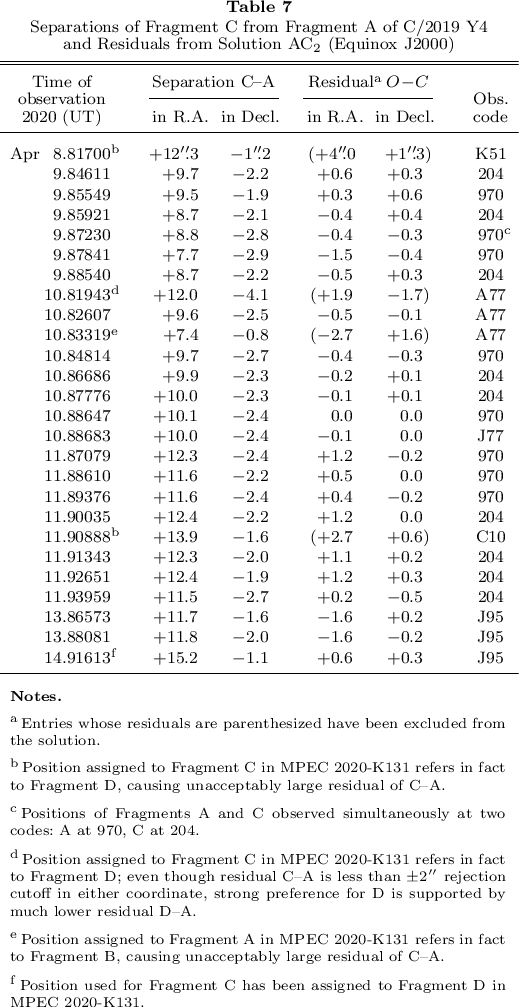}}} %  TABLE 7
\vspace{0.75cm}
\end{table}

Contrary to Solutions BC$_1$ and BC$_2$, Solutions AC$_1$ and AC$_2$
differ from each other considerably.  As seen from both the residuals in
Table~7 and from the plot of the separation trajectory in Figure~5, the
fit to the data by Solution AC$_2$ can hardly be better.  On the other
hand, Solution AC$_1$ has a number of weaknesses.  In the adopted overall
fragmentation scenario it is in fact meaningless, because it implies that
Fragment~C had broken off from Fragment~A {\it before\/} A separated from
B (cf.\ Section~3.1).  In addition, as seen from Table~8, this solution
offers acceptable residuals (below 2$^{\prime\prime}$) from only 20 of the
22~data points and strong systematic trends of about 1$^{\prime\prime}$
are clealy apparent in the distribution of residuals in right ascension.

\begin{figure}[t]
\vspace{0.22cm}
\hspace{-0.18cm}
\centerline{
\scalebox{0.63}{
\includegraphics{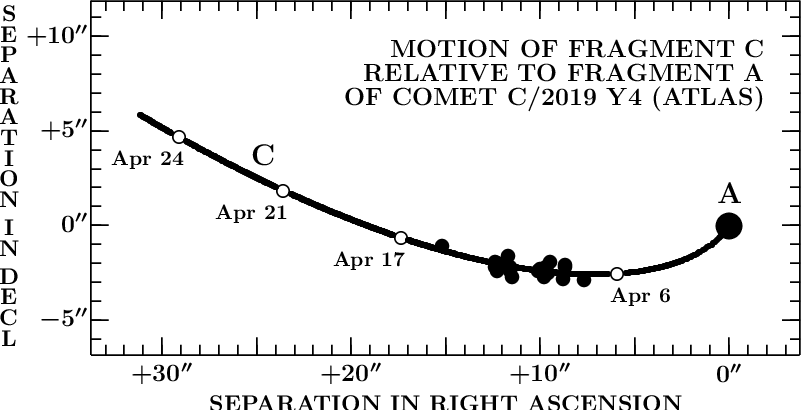}}} %  FIGURE 5
\vspace{0cm}
\caption{Separation trajectory of Fragment C relative to Fragment A
derived from Solution AC$_2$ in Table 5.  The solid circles are the
observations of C from Table 7, while the open circles are ephemeris
positions (at 0 TT) from the model.  Note the excellent agreement
between the data and the model, in sharp contrast to the poor fit in
Figure~4.{\vspace{0.5cm}}}
\end{figure}

The presented evidence suggests rather strongly that it is Fragment~A
that is the parent to Fragment~C.  The failure to get an acceptable
distribution of residuals on the assumption that the parent was the
principal mass B obviously had fundamental roots and was not caused
by inaccurate astrometric observations.

The determined fragmentation sequence has implications for the meaning
of the observed positions of Fragment~C relative to Fragment~B.
Dynamically, their separations represent the sum of separations of
C from A and A from B:
\begin{equation}
{\rm C} \!-\! {\rm B} = ({\rm C} \!-\! {\rm A}) + ({\rm A} \!-\! {\rm B}).
\end{equation}
The averaged daily separations of \mbox{C--B} computed in this fashion
for the mean times of observation of this pair (cf.\ Table~6) are listed
in Table~9, where the residuals $O\!-\!C$ show rather good accord
with the observed separations.

\begin{table}[t]
\vspace{0.21cm}
\hspace{-0.19cm}
\centerline{
\scalebox{0.975}{
\includegraphics{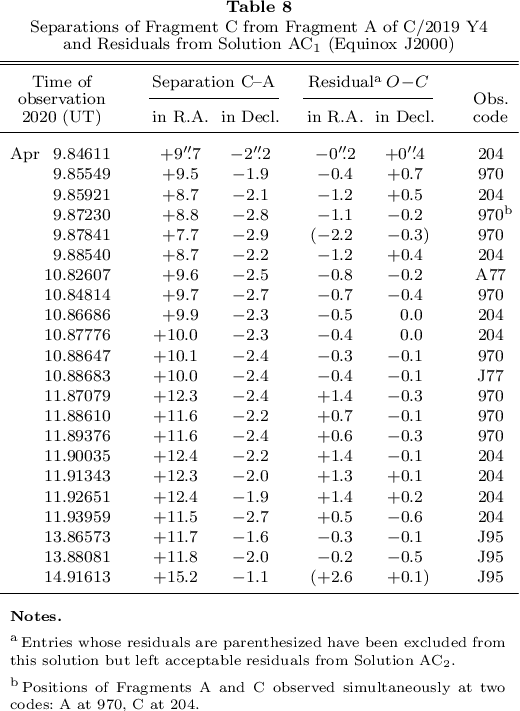}}} %  TABLE 8
\vspace{0.6cm}
\end{table}

\subsection{Fragment D and Its Parent}
The examination of the origin and motion of Fragment~D has followed
closely the routine used in the previous section for Fragment~C,
including the rejection cutoff of 2$^{\prime\prime}$.  Unfortunately,
the identity of fragments in some observations was somewhat uncertain.
Chronologically, the first such case referred to April 10.81943 (code
A77), wherein the MPC assigned Fragment~C had to be replaced with D.
This improved the fit, even though C had been borderline.  A more
difficult was the observation on April 11.90888 (code C10), likewise
requiring that Fragment~C be changed to D, mainly because of the poor
fit in the \mbox{C--A} separation.  Although \mbox{C--B} and \mbox{D--B}
fitted reasonably well, \mbox{C--B} had to be rejected.  The observation
of April 14.86279 (code 970) was a relatively straightforward case;
the position of Fragment~B apparently was in error, as the separation
\mbox{D--A} fitted  well, but \mbox{A--B} and \mbox{D--B} did not at
all.  On the other hand, the observation of April 14.91613 (code J95)
was a major conundrum.  The separations \mbox{C--A} and \mbox{D--A}
pointed to the need to correct the MPC assigned Fragment~D to C, as
\mbox{C--A} fitted well but \mbox{D--A} did not.  But when I tested
the separations \mbox{C--B} and \mbox{D--B}, C was a problem.  The
separation \mbox{A--B} was acceptable, so it was not clear where the
culprit was.  Fragment C appeared preferable to D, but the contradiction
persisted.  The final case was the observation of April 14.91977 (code
Z80), where I also replaced Fragment~D with C, as the separation \mbox{C--B}
offered a better fit than \mbox{D--B}, even though both were acceptable.

The results for Fragment D show similarities with the results for
Fragment~C, except that the numbers of available (B,\,D) and
(A,\,D)~pairs have now been comparable.  The fragmentation models
are summarized in Table~10, which suggests --- like Table~5 for
Fragment~C --- two similar solutions for the motion of D relative
to B and two very different solutions relative to A.

\begin{table}[b]
\vspace{0.6cm}
\hspace{-0.24cm}
\centerline{
\scalebox{0.98}{
\includegraphics{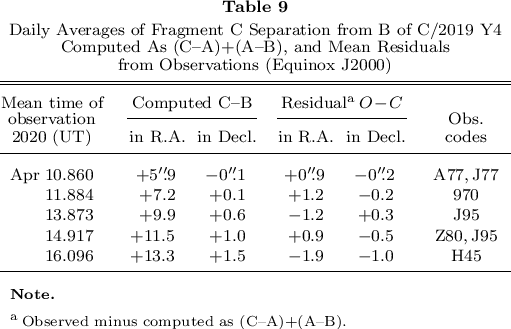}}} %   TABLE 9
\vspace{-0.05cm}
\end{table}
\begin{table*}[t]
\vspace{0.2cm}
\hspace{-0.2cm}
\centerline{
\scalebox{1}{
\includegraphics{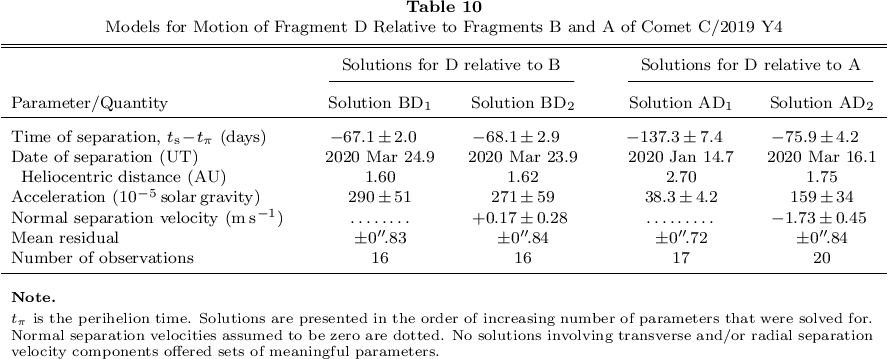}}} %  TABLE 10
\vspace{0.7cm}
\end{table*}

Solution BD$_2$ again includes a normal component of the separation
velocity that is not well determined, so that Solution~BD$_1$ is
preferred.  Unlike in the case of Solution BC$_1$, the distribution
of residuals offered by Solution~BD$_1$ is rather acceptable, as
seen from Table~11 and Figure~6.

\begin{table}[b]
\vspace{0.6cm}
\hspace{-0.16cm}
\centerline{
\scalebox{0.97}{
\includegraphics{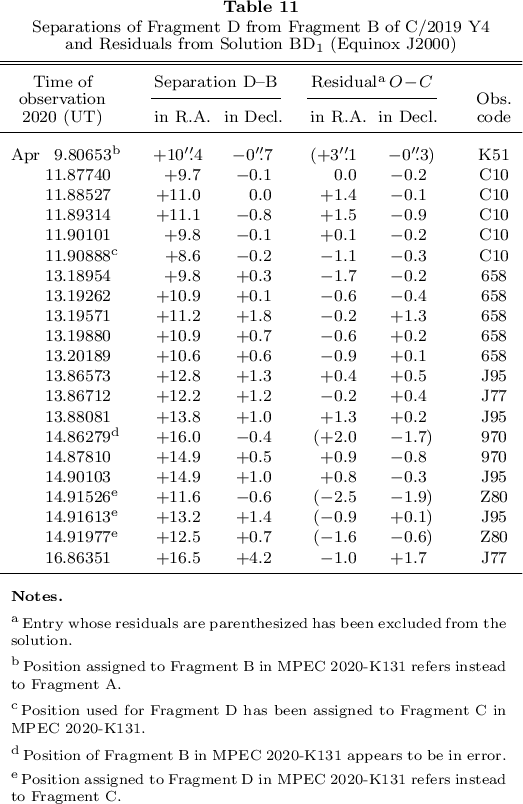}}} %  TABLE 11
\vspace{-0.06cm}
\end{table}

Of the two solutions for the motion of Fragment~D relative to A,
Solution~AD$_1$ is inferior to Solution AD$_2$ in that it has been
unable to fit three of the 20~data points within the rejection cutoff.
The full data set, represented by Solution AD$_2$ with a mean residual
of $\pm$0$^{\prime\prime}\!$.84 (Table~10), has been fitted by Solution
AD$_1$ with a mean residual as high as $\pm$0$^{\prime\prime} \!$.99.
However, unlike Solution AC$_1$, Solution AD$_1$ strictly is not
meaningless because its separation from Fragment~A is found to have
taken place within 1$\sigma$ of the time of separation of A from B.
Although the distribution of residuals is not bad, the solution is shown
above to be weak and the derived separation time is not realistic, being
affected by neglect of the normal component of the separation velocity.
On the other hand, Solution~AD$_2$ and the BD solutions are in fairly
close agreement in that D split off from its parent in mid-, or the
second half of, March.  

\begin{figure}[b]
\vspace{0.7cm}
\hspace{-0.2cm}
\centerline{
\scalebox{0.633}{
\includegraphics{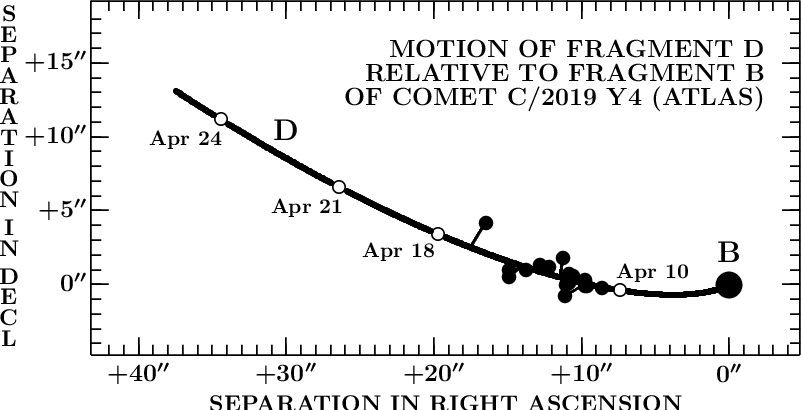}}} %   FIGURE 6
\vspace{0cm}
\caption{Separation trajectory of Fragment D relative to Fragment~B
derived from Solution BD$_1$ in Table~10.  The solid circles are the
observations of D from Table~11, while the open circles are
ephemeris positions (at 0 TT) from the model.{\vspace{-0.06cm}}}
\end{figure}

For comparison with Solution BD$_1$, the distribution of residuals
offered by Solution AD$_2$ is displayed in Table~12 and the separation
trajectory of Fragment~D relative to A is plotted in Figure~7.
The plot indicates that except for three data points at the very
beginning of the observed segment of the trajectory, the fit is in
fact better than the fit to the motion of Fragment~D relative
to B exhibited in Figure~6.

The other point to argue is that it is unlikely that both fragments,
A and B, should have broken up almost simultaneously, one releasing
C, the other D.  On the other hand, it should not be surprising that
the gradually disintegrating A produced, next to the debris, two
subfragments in a row, or perhaps one that almost immediately split
into two.  In any case, there is some evidence for preferring A to B
as the parent to D, but it is tenuous.

\begin{table}[t]
\vspace{0.12cm}
\hspace{-0.21cm}
\centerline{
\scalebox{0.97}{
\includegraphics{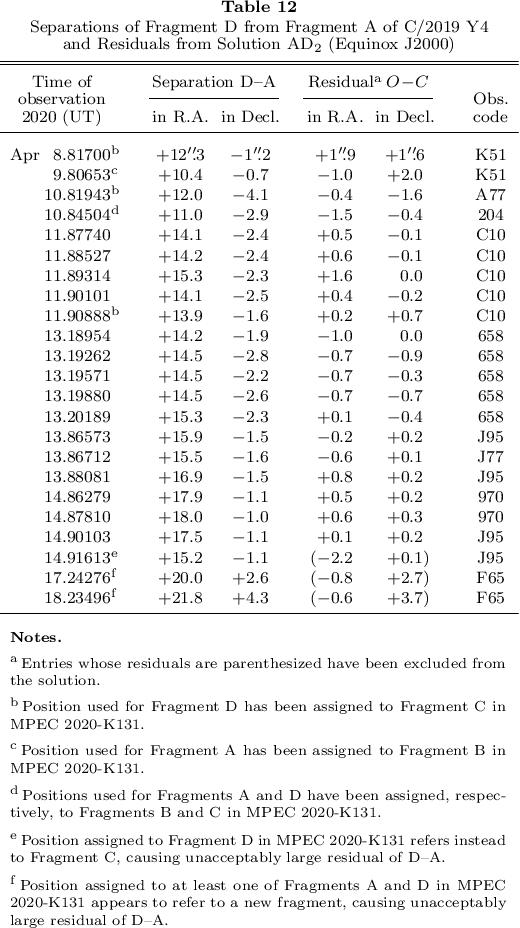}}} %  TABLE 12
\vspace{0.6cm}
\end{table}

\subsection{New Fragment E}
As is apparent from Table 12, the separations of \mbox{(D--A)} on April
17.24276 and 18.23496~UT (code F65) left residuals in declination of
almost 3$^{\prime\prime}$ and 4$^{\prime\prime}$, respectively, from
Solution~AD$_2$.  On the other hand, the separation of \mbox{(A--B)}
left on the 17th an excellent residual in either coordinate from
Solution BA$_1$ (Table~4), indicating that the problem was not with
Fragment~A.  Under these circumstances, one can rule out that the
third fragment on the 17th and the second fragment on the 18th were D.
Since Fragment~C replacing D would have fitted the separation trajectory
even worse, the odds were that on the two days one had to do with a new
fragment, E.

\begin{figure}[t]
\vspace{0.15cm}
\hspace{-0.2cm}
\centerline{
\scalebox{0.64}{
\includegraphics{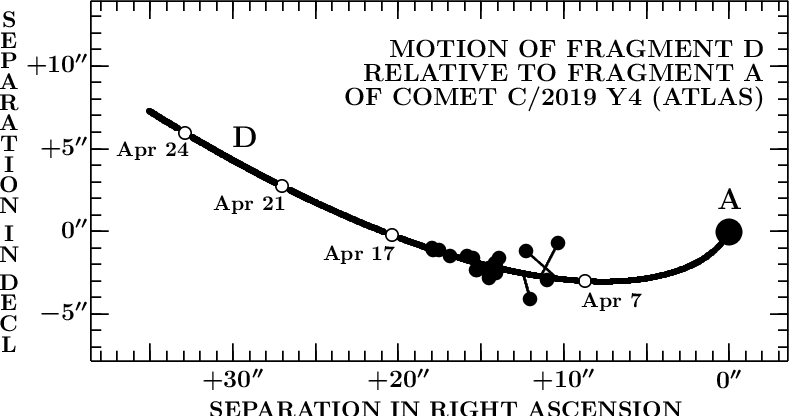}}} %  FIGURE 7
\vspace{0cm}
\caption{Separation trajectory of Fragment D relative to Fragment A
derived from Solution AD$_2$ in Table 10.  The solid circles are the
observations of D from Table 12, while the open circles are ephemeris
positions (0 TT) from the model.{\vspace{0.65cm}}}
\end{figure}

Furthermore, given that the separation of \mbox{(D--A)} on April
9.80653~UT (code K51) left, from Solution~AD$_2$ in Table~12, a large
residual in declination that essentially equaled the rejection cutoff,
it is reasonable to suspect that here too the involved fragment
may have been E rather than D.  Computer runs linking the observations
from April~9.8, 17.2, and 18.2 did indeed lead to successful solutions
AE$_1$ and AE$_2$, displayed in Table~13.  The residuals left by the
latter solution are listed in Table~14.  On account of the extremely
small number of observations, either solution is deemed provisional,
the second being slightly preferable.

The deletion of the \mbox{(D--A)} separation on April~9.8 from the set in
Table~12 had a relatively minor effect on the fragmentation parameters,
leading to Solution AD$_3$ also presented in Table~13; the revised list
of residuals is in Table~15.  It is noted that the removal of the data
point from the 9th had, on the average, a beneficial effect on the
residuals from other observations, the one on the 8th in particular.
In the following I somewhat reluctantly adopt Fragment~A as the
most likely parent of D and Solution~AD$_3$ as the latter's best
available fragmentation solution.

The reason for investigating only the scenario in which Fragment~E
derived from A was the availability of the \mbox{(E--A)} separations on
three days.  This data set, however humble, compares to a single occasion,
on April 17, of the simultaneously observed Fragments B and E.  The
parent of Fragment~E remains unclear and it is merely for the reasons
already commented on in Section~3.3 that one can speculate that the
most likely candidate for being its parent is Fragment~A.

\begin{table*}[t]
\vspace{0.2cm}
\hspace{-0.2cm}
\centerline{
\scalebox{1}{
\includegraphics{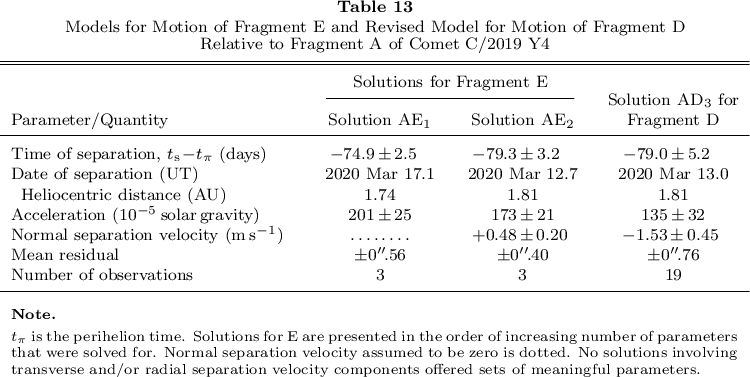}}} %  TABLE 13
\vspace{0.7cm}
\end{table*}

\subsection{Relationship Among Fragments C, D, and E}
The number of simultaneously observed \mbox{(C,\,D)} pairs is very
limited.  From the data in Tables~1 and 2, their total number is
assessed at six.  I have attempted a fragmentation solution to check
whether Fragment~D could have separated from Fragment C rather than
from A (or B).  The result, presented as Solution~CD$_1$ in Table~16,
is inconclusive.  The acceleration in particular is determined with
large uncertainty and no separation velocity could be detected.  As shown
in Table 17, two of the six observations could not be accommodated by
the solution.  There is no evidence that would point at a close
relationship between Fragments~C and D and at their separation from
the parent as a single body.

Similar examinations of possible commonalities between Fragments~C and
E on the one hand and between Fragments~D and E on the other hand
cannot be undertaken because of the absence of reported simultaneous
observations of either pair of these objects.

\begin{table}[b]
\vspace{0.7cm}
\hspace{-0.21cm}
\centerline{
\scalebox{0.97}{
\includegraphics{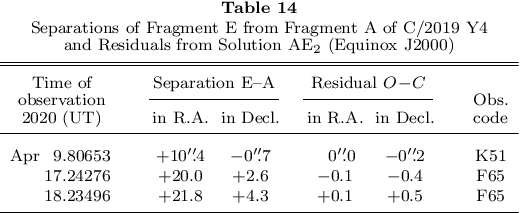}}} %   TABLE 14
\vspace{0cm}
\end{table}

\section{Nuclear Fragmentation of C/2019 Y4:\\Summary and Discussion}
%  SECTION 4
%
% The status of C/2019 Y4, as a small fragment of a very massive comet
% in a long-period orbit (thousands of years), relative to its ``big
% brother'' C/1844~Y1 (Great Comet) parallels that of C/1996~Q1 (Tabur),
% C/2015~F3 (SWAN), or C/2019~Y1 (ATLAS) relative to their intrinsically
% bright sibling C/1988~A1 (Liller).  Another such pair is C/1988~J1
% (Shoemaker-Holt; a minor fragment) vs C/1988~F1 (Levy; a major fragment).
% And a potential intriguing analogy is the enigmatic bright object seen
% with the naked eyes near the Sun by two highly respected astronomers in
% early August 1927 as a minor fragment of C/1847~C1
%
As pointed out in Section~1, the disintegration of comet C/2019~Y4 has
not been entirely unexpected.  Yet the process of its manifestation was
rather unusual.  I have remarked in Section~2 that the shape of the
early portion of the nuclear-condensation's light curve looked very
untypical.  Until late January the {\it apparent\/} magnitude was
constant, which, given that the comet was approaching both the Sun
and Earth, meant that the {\it intrinsic\/} brightness was decreasing
with time.  The average rate of decline between late December and late
January was actually rather steep, varying with heliocentric distance
$r$ approximately as $r^{4.5}$.  Without the knowledge of the subsequent
developments one could have guessed that the comet began to fade out
already near 3~AU preperihelion.  Since this clearly was not the case,
the alternative and still potentially valid scenario could be that of
the comet's light curve following a subsiding branch of an earlier
outburst that took place prior to mid-December.

Under these circumstances, the comet's sudden and sustained brightening
starting in late January looked as if the comet got a second wind.
And since it is fairly common that an episode of nuclear splitting
is accompanied by an outburst (e.g., Sekanina 2010), a prolonged
brightening should, by extension, accompany a process of continual
(recurring in rapid succession) or essentially continuous
(uninterrupted) fragmentation over an extended period of time.  If
the particle velocity was rather low, new debris
got into the inner coma before much of earlier one managed to leave it.
Also, once the process started, the increasing energy input from the
Sun, as the comet kept approaching it, did facilitate debris release
at ever higher rates.

\begin{table}[b]
\vspace{0.7cm}
\hspace{-0.21cm}
\centerline{
\scalebox{0.97}{
\includegraphics{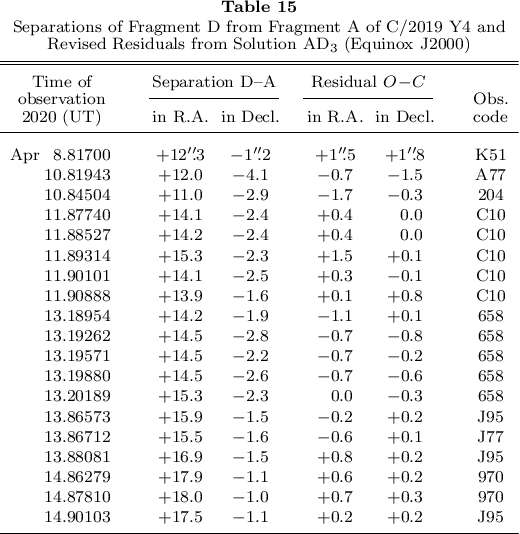}}} %  TABLE 15
\vspace{-0.02cm}
\end{table}

This mode of fragmentation makes it difficult to model, because the
resulting normalized nongravitational acceleration on the fragmenting
object increases systematically with time at a high rate, as the object
rapidly contracts.  Application of a standard model that does not allow
for this effect returns thus a value that represents a crude average
over the relevant period of time.

\begin{table}[t]
\vspace{0.14cm}
\hspace{-0.2cm}
\centerline{
\scalebox{1}{
\includegraphics{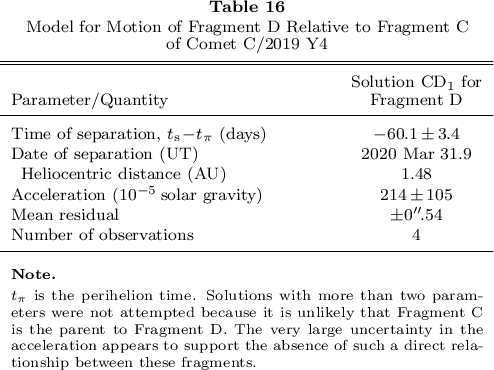}}} %   TABLE 16
\vspace{0.7cm}
\end{table}

\subsection{Fragmentation Sequence}
If the late January sudden change in the light curve of the nuclear
condensation signaled a fragmentation event, it unquestionably was a
breakup of the parent mass into Fragments A and B.  Although the
model failed to converge when the time of separation became one of
the parameters to solve for, the distribution of residuals from the
33~used separations presented in Table~4 was deteriorating dramatically
as the separation time was being moved forward.  While the mean
residual reached $\pm$0$^{\prime\prime}\!$.522 for a separation time of
130~days before perihelion (from adopted Solution~BA$_1$ in Table~3),
it increased to $\pm$0$^{\prime\prime}\!$.528 at 100~days before
perihelion and to $\pm$0$^{\prime\prime}\!$.563 at 70~days before
perihelion.  A formal minimum of $\pm$0$^{\prime\prime}\!$.518 was
reached 650~days before perihelion, an unrealistic solution.

Of the other fragments, the conclusion from the computations in the
preceding sections is that separation from Fragment~A was very
probable for C, perhaps likely for D, and possible for E.  Adopting
for the three objects the fragmentation solutions of, respectively,
AC$_2$ from Table~5, AD$_3$ from Table~13, and AE$_2$ from Table~13,
one finds that D and E broke off essentially simultaneouly, while C
about a week later, but the C and D events were within 1$\sigma$ of
each other.  These separation dates were centered on mid-March and
approximately coincided with the sudden drop in the comet's {\it
rate\/} of brightening, days before the light curves in Table~2
reached the peak.

\begin{table}[b]
\vspace{0.8cm}
\hspace{-0.21cm}
\centerline{
\scalebox{0.97}{
\includegraphics{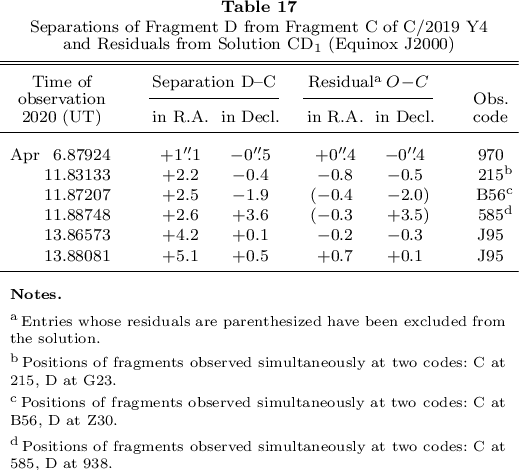}}} %  TABLE 17
\vspace{0cm}
\end{table}

\subsection{Fragments' Detection, Activity, and\\Separation Trajectories}
Given that the duplicity of the comet's nucleus~was~not detected until
early April, the breakup in late January meant that one component ---
the principal mass B, as it turned out --- remained unobserved for at least
11~weeks.  The disintegrating component~A was the only~part of the original
nucleus --- itself a companion --- that~was under observation continuously
from the breakup until late April.  In addition, the separation
trajectory of Fragment~A relative to B in Figure~3 suggests that in
mid-March, when only A was seen, the two fragments were nearly
8$^{\prime\prime}$ apart, while by mid-April, when both A and B were
observed, their separation diminished to less than 5$^{\prime\prime}$.
The separation distance had obviously no effect on the detection
of the duplicity.  Instead, it was low activity of Fragment~B prior
to April that was keeping its brightness below the detection threshold
of the telescopes used.  It is likely that the HST would have been
able to image the duplicity for most of the time it lasted, beginning
not later than early February.

Interestingly, Fragments C, D, and E were all observed within four
weeks or so of their presumed release from Fragment~A, starting at
separations of 10$^{\prime\prime}$ to 12$^{\prime\prime}$.  

\subsection{Fragments' Nongravitational Accelerations,\\Lifetimes, and
 Relative Sizes}
The sublimation-driven nongravitational acceleration $\gamma$ is an
important fragmentation parameter, because it is determined by the
physical properties of the fragment, just as in Whipple's (1950) model the
nongravitational acceleration of a comet is determined by its physical
properties.  However, when the fragment's motion is measured relative
to the parent nucleus, the derived value is a {\it differential\/}
acceleration.  The absolute acceleration of the fragment is the sum
of its differential acceleration~and the parent's acceleration.

The acceleration is statistically correlated with the fragment's
lifespan, which is conveniently measured by the endurance, $\cal E$,
defined as the difference between the separation time and
the time of last observation, reduced to a normalized heliocentric
distance of 1~AU, and expressed in equivalent days (or e-days).  For
comets in nearly-parabolic orbits (Sekanina 1982)
\begin{equation}
{\cal E} = 40.5 \, q^{-\frac{1}{2}} \! \left(u_{\rm fin} \!-\!
 u_{\rm frg}\right),
\end{equation}
where $q$ is the perihelion distance (in AU) and $u_{\rm frg}$ and $u_{\rm
fin}$ are the true anomalies (in radians) at the times of separation
and final observation, respectively.  Because of effects
due to observing conditions, the measured lifespan is in
general a lower limit to the true lifespan.

The limited statistics of two dozen data points suggest the existence of
three basic categories of cometary fragments:\ {\it persistent\/}, {\it
short-lived\/}, and {\it minor\/} (Sekanina 1982).  As already noted in
Section~1, the endurance was found to vary with the acceleration as
\begin{equation}
{\cal E} = C \, \gamma^{-0.4},
\end{equation}
where \mbox{$C = 800$ e-days} for the most massive, persistent
fragments, 200~e-days for the short-lived fragments, and
87~e-days for the minor fragments.

From Solutions BA$_1$, AC$_2$, AD$_3$, and AE$_2$~one~finds,~in units
of 10$^{-5}$\,the solar gravitational acceleration,~the nongravitational
accelerations of 21.7\,$\pm$\,2.8 for~\mbox{Fragment} A relative to B and
161\,$\pm$\,30, 135\,$\pm$\,32, and 173\,$\pm$\,21, respectively, for
Fragments~C, D, and E relative to A.  Fragment~A is positively a
short-lived fragment, whose acceleration implies an endurance of
58\,$\pm$\,3~\mbox{e-days}. The other fragments are apparently short lived
as well, in which case the endurances equal 26.2\,$\pm$\,2.0~\mbox{e-days}
for Fragment~C, 28.1\,$\pm$\,2.7~\mbox{e-days} for D, and
25.5\,$\pm$\,1.2~\mbox{e-days} for E.  Should they be minor fragments,
their endurances would be too short:\ 10.8\,$\pm$\,0.8~\mbox{e-days},
11.5\,$\pm$\,1.1~\mbox{e-days}, and 10.6\,$\pm$\,0.5~\mbox{e-days},
respectively.

These numbers apply on the assumption that the principal mass B was
affected by no nongravitational acceleration.  As the comet was itself
a companion fragment of a larger comet, this is unlikely to be the
case.  Indeed, the orbit computed by the MPC Staff (2020) indicates
that between 20~December 2019 and 3~April 2020 the comet's motion was
affected by a strong nongravitational acceleration.  Of course, if
the nucleus split into Fragments~A and B around 22~January, the MPC
elements timewise cover only 31~percent of the motion of the pre-split
nucleus and 69~percent of the motion of Fragment~A, as Fragment~B
was not seen until after 3~April.  As the MPC Staff determined the
orbit using Marsden et al.'s (1973) model with the standard $g(r)$
nongravitational law, it is necessary to convert the integrated effect
of the radial sublimation-driven force from the Marsden et al.\
nomenclature to the style used in this paper.

The fundamental condition to be satisfied is~that~the dominant
nongravitational effect in the radial direction, whose magnitude in
the period from 20~December to 3~April was measured in the MPC Staff's
{\vspace{-0.035cm}}(2020) orbital solution by a parameter \mbox{$A_1 =
+16.16 \times \! 10^{-8}$\,AU day$^{-2}$}, be equated with the sum of
the radial nongravitational effects, expressed in the style used in this
paper, (i)~on the pre-split nucleus between 20~December and 22~January,
defined by a parameter $\gamma_{\rm par}$; {\it plus\/} (ii)~on
Fragment~A between 22~January and 3~April.  The latter part
consists of the total radial nongravitaional effect on Fragment~B,
defined by a parameter $\gamma_{\rm b}$; {\it plus\/} the radial
nongravitational effect on Fragment~A relative to Fragment~B, defined
by the tabulated parameter \mbox{$\gamma_{\rm ba} = 21.7$ units}
of 10$^{-5}$\,the solar gravitational acceleration.  The condition
thus reads
\begin{eqnarray}
\int_{t_{\rm beg}}^{t_{\rm end}}  \!\!\! A_1 \, g(r)\,dt & = & \!
 \int_{t_{\rm beg}}^{t_{\rm frg}} \!\!\! \gamma_{\rm par} \, \Gamma\!_\odot
 \! \left(\:\!\! \frac{r_\oplus}{r} \! \right)^2 \!\! dt \nonumber \\
 & + & \!\! \int_{t_{\rm frg}}^{t_{\rm end}} \!\!\! \left( \gamma_{\rm b}
 \! + \! \gamma_{\rm ba} \right) \Gamma\!_\odot \! \left( \:\!\!
 \frac{r_\oplus}{r} \! \right)^2 \!\! dt,
\end{eqnarray}
where $t_{\rm beg}$, $t_{\rm end}$, and $t_{\rm frg}$ are, respectively,
the times of the beginning and the end of the orbital arc, integrated by
the MPC Staff (2020), and the time of fragmentation.  Expressed in days from
perihelion, they are equal to \mbox{$t_{\rm beg} = -163$ days}, \mbox{$t_{\rm
end} = -58$ days}, and, as adopted, \mbox{$t_{\rm frg} = -130$ days};
\mbox{$r_\oplus = 1$ AU}, so that{\vspace{-0.04cm}} $\gamma_{\rm par}$
and $\gamma_{\rm b}$ refer to 1~AU and, like $\gamma_{\rm ba}$, are in units
of 10$^{-5}$ the solar gravitational acceleration; at 1~AU from the Sun
this~unit~equals $\Gamma\!_\odot = 0.593 \times \! 10^{-5}$\,cm s$^{-2}
= 0.296 \times \!10^{-8}$\,AU day$^{-2}$.  {\vspace{-0.03cm}}Employing
Kepler's Second Law and elementary operations, and leaving $r_\oplus$ out,
Equation~(11) can be rewritten thus:
\begin{equation}
A_1 \! \Im_{\rm beg,end} \!=\:\!\! \Gamma\!_\odot \! \left[ \gamma_{\rm par}
 (u_{\rm frg} \!-\! u_{\rm beg}) \!+\! (\gamma_{\rm b} \!+\! \gamma_{\rm ba})
 (u_{\rm end} \!-\! u_{\rm frg}) \right]\!,
\end{equation}
where $u_{\rm beg}$, $u_{\rm end}$, and $u_{\rm frg}$ are the true anomalies
(in~radians) at, respectively, $t_{\rm beg}$, $t_{\rm end}$, and $t_{\rm
frg}$; and
\begin{equation}
\Im_{\rm beg,end} = \!\! \int_{u_{\rm beg}}^{u_{\rm end}} \!\!\!r^2 g(r)
 \, du = \!\! \int_{r_{\rm end}}^{r_{\rm beg}} \!\!\!r \! \left( \!
 \frac{r}{q} \!-\! 1 \!\! \right)^{\!\!-\frac{1}{2}} \!\!\! g(r) \, dr.
\end{equation}
The integration over $r$, with \mbox{$r_x = r(t_x)$}, is a parabolic
approximation valid on the assumption that \mbox{$u_{\rm beg} < 0$} and
\mbox{$u_{\rm end} < 0$}.

The condition (12) includes two unknowns:\ $\gamma_{\rm par}$ and $\gamma_{\rm
b}$.  Its solution requires an additional condition~on~the two
accelerations.  If the parent and Fragment~B were about equally active,
one would expect that \mbox{$\gamma_{\rm par} < \gamma_{\rm b}$}, the
two values the closer to each other the smaller Fragment~A relative
to B.  Because Fragment~A was much more active than B, it is likely
that in relative terms the parent was more active than Fragment~B,
in which case $\gamma_{\rm par}$ could equal $\gamma_{\rm b}$ or even
exceed it.  As a first-guess estimate I assume equality, in which case
condition (12) becomes
\begin{equation}
\gamma_{\rm par} = \gamma_{\rm b} = \frac{3.38 \times \! 10^8 A_1
 \Im_{\rm beg,end} - \gamma_{\rm ba} (u_{\rm end} \!-\! u_{\rm
 frg})}{u_{\rm end} \!-\! u_{\rm beg}}.
\end{equation}
For the considered MPC orbit, \mbox{$u_{\rm end} \!-\! u_{\rm beg} =
0.2813$}~radian, \mbox{$u_{\rm end} \!-\! u_{\rm frg} = 0.2304$}~radian,
and from Equation~(13) \mbox{$\Im_{\rm beg,end} = 0.1167$ AU$^2$}, so that
\begin{equation}
\gamma_{\rm par} = \gamma_{\rm b} = 4.9 \; {\rm units\,\,of\,\,10}^{-5} \;
 {\rm solar\,\,grav.\,\,accel}.
\end{equation}
More generally, one could assume
\begin{equation}
\gamma_{\rm b} = g \gamma_{\rm par},
\end{equation}
where \mbox{$g > 0$} and \mbox{$g \neq 1$}.  The solution (14) is now
slightly modified,
\begin{equation}
\gamma_{\rm par} = \frac{3.38 \times \! 10^8 A_1 \Im_{\rm beg,end} -
 \gamma_{\rm ba}(u_{\rm end} \!-\! u_{\rm frg})}{(u_{\rm end} \!-\!
 u_{\rm beg}) + (g \!-\! 1) (u_{\rm end} \!-\! u_{\rm frg})}.
\end{equation}
One can show that if \mbox{$\gamma_{\rm par} = \gamma_{\rm b} > 0$},
then \mbox{$\gamma_{\rm par} > 0$} and \mbox{$\gamma_{\rm b} > 0$}
for any \mbox{$g > 0$}.  The only condition for \mbox{$\gamma_{\rm
par} > 0$} is a constraint on the ratio of $\gamma_{\rm ba}$ to $A_1$,
\begin{equation}
\frac{\gamma_{\rm ba}}{A_1} < \frac{3.38 \times \! 10^8 \,
 \Im_{\rm beg,end}}{u_{\rm end} \!-\! u_{\rm frg}},
\end{equation}
where the units are, as above, 10$^{-5}$\,the solar gravitational
acceleration for $\gamma_{\rm ba}$ and AU~day$^{-2}$ for $A_1$.

Limited experimentation with the factor $g$ suggests that for values
moderately exceeding, or slightly lower than, unity, the{\vspace{-0.05cm}}
accelerations $\gamma_{\rm par}$ and $\gamma_{\rm b}$ do not differ
from (15) by more than a few units of 10$^{-5}$\,the solar gravitational
acceleration.  As seen, $\gamma_{\rm b}$ is about a factor of 4 to 5
smaller than $\gamma_{\rm ba}$ and does not amount to more than a few
percent (well within the uncertainty) of the high nongravitational
accelerations that Fragments~C, D, and E are subjected to.

\begin{table*}[t]
\vspace{0.18cm}
\hspace{-0.1cm}
\centerline{
\scalebox{1}{
\includegraphics{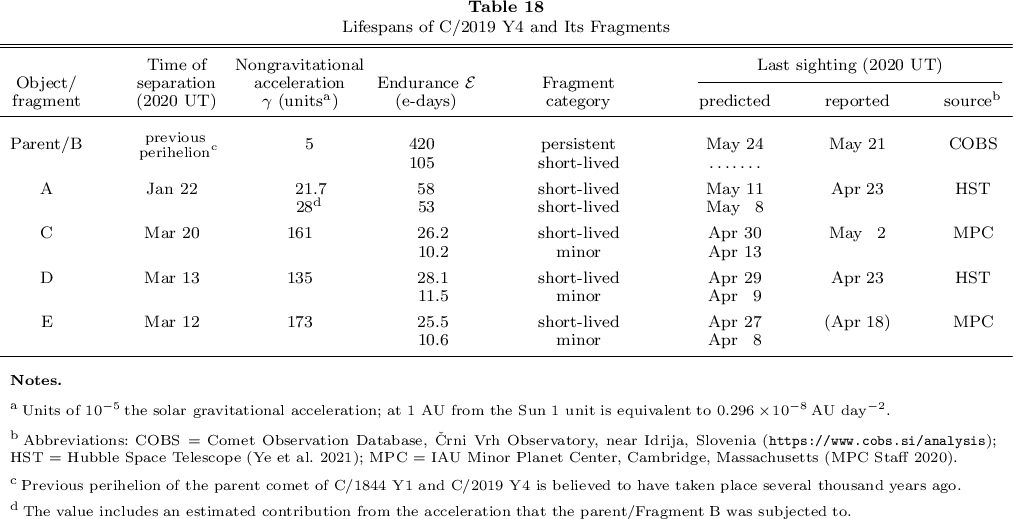}}} %  TABLE 18
\vspace{0.6cm}
\end{table*}

Turning back to the lifespan of the fragments, I note that by eliminating
the endurance from Equations~(9) and (10) it is possible to approximately
predict the time of final observation of a fragment, $t_{\rm fin}$, from
the model parameters --- the time of separation, $t_{\rm frg}$, and the
nongravitational acceleration, $\gamma$ --- when the fragment's category
(persistent, short-lived, or minor), determining the parameter $C$, is
known.  The prediction can be tested on the data available from
observations.  In a parabolic approximation, adequate for this purpose,
the procedure includes three steps as follows:
\begin{eqnarray}
u_{\rm fin} & = & u_{\rm frg} + 0.0247 \, C q^{\frac{1}{2}}
 \gamma^{-\frac{2}{5}} \!, \nonumber \\[0.1cm]
r_{\rm fin} & = & \frac{q}{\cos^2 \frac{1}{2} u_{\rm fin}}, \nonumber \\[0.1cm]
t_{\rm fin} \!-\!t_\pi & = & \mp 27.404 \, (r_{\rm fin} \!+\! 2q)
 (r_{\rm fin} \!-\! q)^{\frac{1}{2}} \! ,
\end{eqnarray}
where $t_\pi$ is the time of perihelion.  As before, the true anomalies
are in radians, the distances in AU, and the times in days, the signs
giving $t_{\rm fin}$ before/after perihelion.

The predicted dates of last sighting are compared in Table 18 with the
reported times.  The agreement~is~fairly good, when appropriate fragment
categories are assigned.  This is dramatically seen in the case of
the principal fragment~B, the most massive part of the pre-split comet,
a companion to the Great Comet of 1844, as already~noted.  Let us assume
that the two comets separated from their parent at the time of previous
perihelion, several thousand years ago.  If C/2019~Y4 should have
survived~as~a persistent fragment until the 2020 perihelion, its endurance
would have been 506~\mbox{e-days}.  From $\gamma_{\rm par}$ in Equation~(15)
its endurace is expected to equal 420~\mbox{e-days}, last observed just
86~\mbox{e-days} short of the full revolution, a period of time equivalent
to $\sim$7~days.  Since the endurance varies as time weighted{\vspace{-0.04cm}}
by $r^{-2}$, the ratio 86/7\,$\simeq$\,12.3 is, as it should be, about halfway
between (1/0.34)$^2 \simeq 8.7$, where 0.34~AU was the heliocentric distance
on May~24, and (1/0.25)$^2$\,=\,16, where 0.25~AU was the perihelion distance
on May~31.  If the comet were instead a short-lived fragment, its endurance
would have been only a little more than 100~\mbox{e-days} and it would
have fragmented and disintegrated at some point after the previous
perihelion, on its way to, and long before reaching, aphelion. C/2019~Y4
would never have been discovered.

By contrast, Fragment A was a short-lived fragment, which should have
lasted until early to nearly mid-May.  Of the few data sources that
I inspected, it was the last time seen in the images taken with
the HST on 23~April and its lifespan must have been longer
than that.  The three fragments that separated, presumably from A,
around mid-March, were obviously also~short~lived.  They
could not have been minor fragments, given that they were observed
after \mbox{8--13}~April.

Even though the definition of the {\it last sighting\/} is not
straightforward, the introduction of the endurance is very beneficial
in that it offers some insight into the morphology of comet
fragmentation products and demonstrates that the sublimation-driven
nongravitational acceleration has important physical implications
in more than one respect.

The disintegration time of a sizable fragment differs from the
disintegration time of a comet as such.  The laws of cascading
fragmentation imply that a comet's particulate debris survives and
is detectable as a fairly condensed cloud over a period distinctly
longer than the individual massive fragments because the debris
has a much higher ratio of cross-sectional area to mass.  There is
nothing controversial, especially among comets of small perihelion
distance, about large fragments dissolving before, while the comet's
remains after, perihelion.  Indeed, C/2019~Y4 was reported to have
been detected by the HI-1 imager of STEREO~A as late as 6~June (Knight
\& Battams 2020), about a week following perihelion.

\begin{figure*}
\vspace{0.17cm}
\hspace{-0.1cm}
\centerline{
\scalebox{0.74}{
\includegraphics{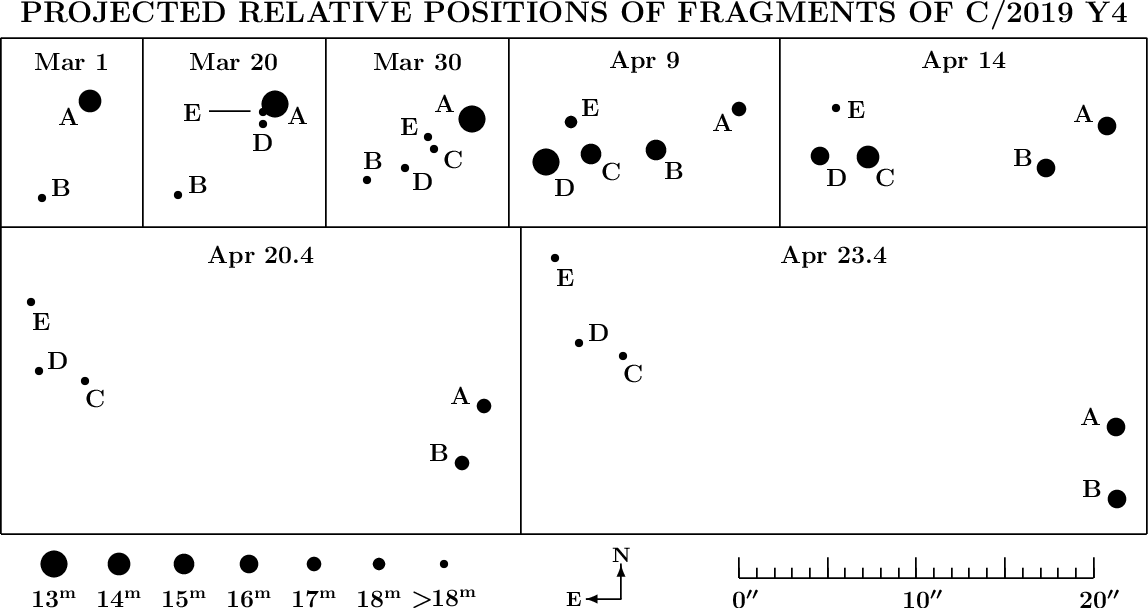}}} %  FIGURE 8
\vspace{-0.3cm}
\caption{Relative positions of the fragments of comet C/2019~Y4 on
seven dates between 1~March and 23~April 2020.  The times for the
first five dates are 0~UT, for the last two dates they approximately
coincide with the times of the HST imaging.  The positions have been
derived from Solutions BA$_1$, AC$_2$, AD$_3$, and AE$_2$.  The size
of each magnitude symbol, assigned for orientation purposes only, is
an approximate match to an average daily nuclear brightness reported
to the MPC; a fragment that was not reported is usually assumed to
have been too faint to detect (of magnitude $>$18).  The scale and
orientation (the north up, east to the left) are also shown.{\vspace{0.6cm}}}
\end{figure*}

An important property of the nongravitational acceleration of a fragment
is that it serves as a crude size estimate of the object.  The
acceleration varies as the mass-loss rate of sublimated gas from its
surface and inversely as the object's mass.  The mass-loss rate varies
as the square of the size, while the mass as the cube of the size.
Accordingly, the nongravitational acceleration varies inversely as
the size.

One can of course think of variations on the basic sublimation model,
including the nuclear shape, rotation, etc.  For example, the following
empirical relation approximates a link between the acceleration $\gamma$
(again{\vspace{-0.03cm}} in units of 10$^{-5}$\,the solar gravitational
acceleration) and the fragment's initial diameter $\cal D$ (in km)
implied by a slightly more complex model (Sekanina 1982):
\begin{equation}
{\cal D} = 2.16 \, \gamma^{-0.92}.
\end{equation}
For the pre-split nucleus of C/2019~Y4 this law gives an effective
initial diameter of 500~meters, for Fragment~A about 100~meters,
and for Fragments~C, D, and E between \mbox{16--20}~meters.

\subsection{Projected Relative Positions of Fragments\\As Function of Time}
The fragmentation sequence of C/2019~Y4, jointly with the peculiar
separation trajectory of Fragment~A relative to B (Figure~3), are behind
the rather intricate motions of the individual fragments relative to one
another in projection onto the plane of the sky.  The picture is further
complicated by their rapid and irregular brightness variations.

The projected relative positions of the five fragments as a function of
time are illustrated in Figure~8 by their status on seven selected
dates, including the two on which the disintegrating comet was imaged
by the HST.  A particularly confusing situation is seen to have developed
in early April, when the multiplicity was first detected.  The three
more recently born fragments were moving in a general direction of the
line connecting the formerly existing Fragments~A and B, not to mention
that it was at this very time that B suddenly became bright enough to
detect.  Subsequently, the high nongravitational accelerations of C, D,
and E caused the rapid eastward motions of these fragments away from A
and B.

\section{Orbit of the Great Comet of 1844}
The early history of major effort to determine the orbit of comet C/1844~Y1
was a little wobbly.  The first to get involved was Bond (1850), in part
because he wanted to examine the possible identity of this object with
the comet of 1556.  Bond collected 138~astrometric observations of the
comet made between 24~December 1844 and 12~March 1845, organizing them
into seven normal places.  He then selected 1~January 1845 as an
osculation epoch and accounted for the perturbations by the ``principal''
planets.  The result was a set of hyperbolic elements with a perihelion
distance of 0.2517170~AU and an eccentricity of 1.00035303.  Bond did
not compute the standard errors of the orbital elements, but his table
of residuals for the normal places implies a mean residual of
$\pm$4$^{\prime\prime\!}$.4.

Sixty years later, Fayet (1910) reviewed Bond's orbital work on this
comet and discovered a serious mistake in the equations of condition
in declination (or, rather, in the polar distance that Bond actually
worked with).  The net result of Fayet's reexamination was an elliptic
orbit, with a perihelion distance of 0.2508701~AU and an eccentricity
of 0.9997302; again, no errors of the elements were provided.  The
corresponding osculating orbital period was about 28,000~yr.  Fayet
did not compute the original period, but it is highly unlikely that
it would have come out much shorter than the osculating value.  In
addition, Fayet also derived a parabolic solution, whose fit to the
observations was only marginally worse, so slightly in fact that
Marsden (1972) incorporated the parabola into the first edition of
his {\it Catalogue of Cometary Orbits\/}, mentioning the ellipse
only in the notes.

I got interested in the physical behavior of C/1844 Y1 as part of my
extensive investigation of anomalous tails (or antitails) of comets
in 1974 (Section 6).  At the time I wanted to learn whether the
astrometric observations were adequate for determining whether the
comet could be a member of the Oort Cloud and decided to calculate
its orbit from scratch. Unlike Bond and Fayet, I employed only the
most consistent positional data that left residuals not exceeding
$\pm$3$^{\prime\prime}$, which restricted their total number to merely
41, or 30~percent of the number collected by Bond.  The outcome of my
effort was first listed (as unpublished) in the second edition of
\mbox{Marsden's} (1975) {\it Catalogue\/} and eventually
incorporated into the summary paper by Marsden et al.\,(1978).
The solution left a mean residual of $\pm$1$^{\prime\prime}\!$.81,
suggesting the comet's original orbital period of
7600\,$\pm$1100~yr.  While there is no doubt that C/1844~Y1 was
not an Oort Cloud comet, the mean error of about 15~percent of the
orbital period is uncomfortably high for the present study.

More recently,\,S.\,Nakano (Green 2020a) reported~a~new orbital
determination of the comet by T.\ Kobayashi, who used another
set of 41~observations that left a mean residual as high as
$\pm$5$^{\prime\prime}\!$.2.  The osculating orbital period
came out to equal 3720~yr, corresponding to an original orbital
period of 4000~yr, more than 3$\sigma$ shorter than the orbit
I had derived in 1974.  The high mean residual makes the new
orbit suspect and, in any case, the discrepancy between the two
results allows one only to conclude that the comet returned to
perihelion after several thousand years.  The orbital periods
derived for C/2019~Y4 are of little value because of the
complications caused by the major nongravitational effects
(Marsden et al.\ 1973).

\section{Great Comet of 1844 and Its Antitail}
The comet, which suddenly burst into sight as a naked-eye object about
4~days after perihelion, was referred~to either as the Great Comet of
1844 (or 1844/1845)~or~as comet Wilmot.~The last name was based on
the strength of Maclear's (1845a) dubious claim that the comet~was
``first seen by Captain Wilmot,'' subsequently corrected by Maclear
(1845b) himself.

In connection with my research on comet Kohoutek, C/1973~E1 (Sekanina
1973, 1974a; Sekanina \& Miller 1976), I realized that the conditions
for displaying a tail {\it projecting sunward\/} and consisting
exclusively of large particulates (typically $>$100~microns across), could
be {\it predicted\/} for any comet sufficiently rich in dust.  A feature
of this kind is referred to as an {\it anomalous tail\/} or, shortly, an
{\it antitail\/}.  In the subsequent investigations of antitails (e.g.,
Sekanina 1974b, 1976), I determined that C/1844~Y1 was one of the
objects for which favorable conditions to display an antitail extended
over a number of days centered on the time of the Earth's transit across
the comet's orbital plane.  The longitudes of the Earth and the comet's
ascending node coincided on 1845 January~18.6~UT, 5~days before the full
Moon.

\begin{figure}[t]
\vspace{0.17cm}
\hspace{-0.18cm}
\centerline{
\scalebox{0.7}{
\includegraphics{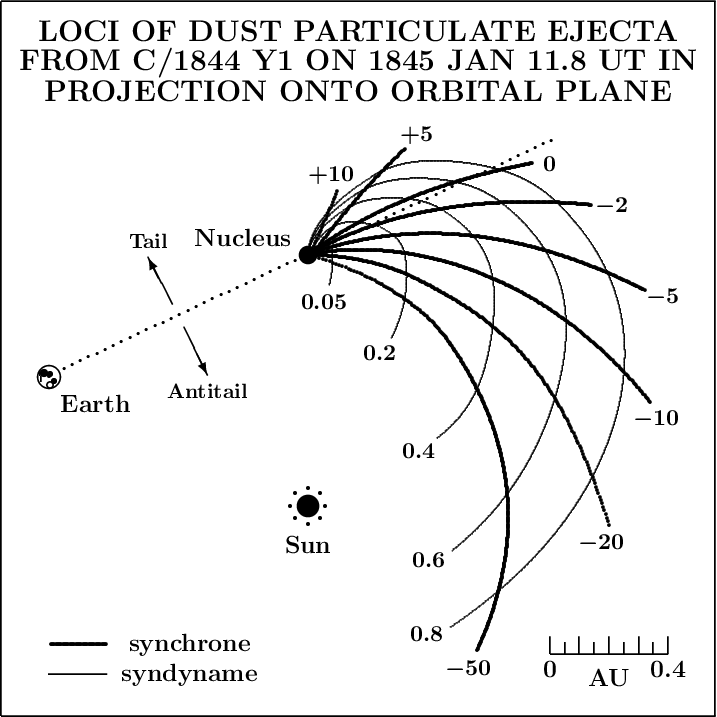}}} %  FIGURE 9
\vspace{0cm}
\caption{Synchrones and syndynames of dust ejected from the nucleus of
comet C/1844~Y1 for the time of observation by T.\ Maclear on 1845
Jan 11.8~UT in projection onto the orbital plane of the comet.  The
synchrones are the thick curves, each marked by the time of ejection
from the nucleus measured in days from perihelion (1844 Dec~14.2~UT).
They cover a period of time from 50~days before perihelion (1844 Oct~25;
$-$50) to 10~days after perihelion (1844 Dec~24; +10).  The age of a
synchrone at the time of observation is derived by subtracting the
marked ejection time from 28.6~days.  The syndynames are the thin
curves, each providing the magnitude of the solar radiation pressure
acceleration $\beta$ affecting the particles and confined in the plot
to a range from 0.05 to 0.8~the solar gravitational acceleration.
Particle size varies inversely as the magnitude of the radiation
pressure effect.  Note that, as seen from the Earth, larger particles
(\mbox{$\beta < 0.2$}) ejected earlier than about 2~days before
perihelion project at the time of observation on the sunward side
of the comet's head as an antitail.  In practice, the visible
antitail consists only of large particles whose \mbox{$\beta < 0.05$}
(Table~19 and the text).  Also note that microscopic grains
(\mbox{$\beta > 0.6$}) released from the comet before late
October 1844 were at the time of observation located on the other
side of the Sun from the comet.  These grains were scattered too
widely in space to be detected.  The Earth was at the time 0.085~AU
above, and moving towards, the comet's orbital plane, transiting it
7~days later.{\vspace{0.6cm}}}
\end{figure}

I was able to locate two independent references to the antitail's
detection in the relevant period of time.  One was a report by
Maclear (1845b), Cape Observatory, who described the appearance
of the comet on January~11:\
%
% , for which time the distribution of
% dust ejecta in the comet's orbital plane is plotted in Figure~9:\
%
{\sl ``[A] faint ray of luminous matter, about 1$\frac{1}{4}$
degree in length, was seen to extend from the anterior portion
of the comet's head in a direction opposite to that of the tail.
The breadth of this ray near the head was about~2$^{\,\prime\!}$,
increasing slightly towards the extremity.  Its borders were
comparatively well defined, and the light gradually diminished
in intensity from that portion nearest the comet's head, until
it became insensible.''}

Maclear then continued:\ {\sl ``From the 18th to the 27th, the tail
and the anterior ray were both rendered invisible by the moonlight.
Viewed with the comet-sweeper on the latter evening, the northern
border of the \ldots anterior ray appeared distinct and sharply
defined, but [its] southern border could no longer be traced. \ldots
Viewed with the 46-inch achromatic, \ldots the  anterior extension of
the light could be traced as a distinct ray for about 5$^{\,\prime}$
from the comet's head, the luminous matter on the southern border
becoming then diffused and scattered, while the northern border
continued well defined.''}

\begin{table*}[t]
\vspace{0.17cm}
\hspace{-0.15cm}
\centerline{
\scalebox{1}{
\includegraphics{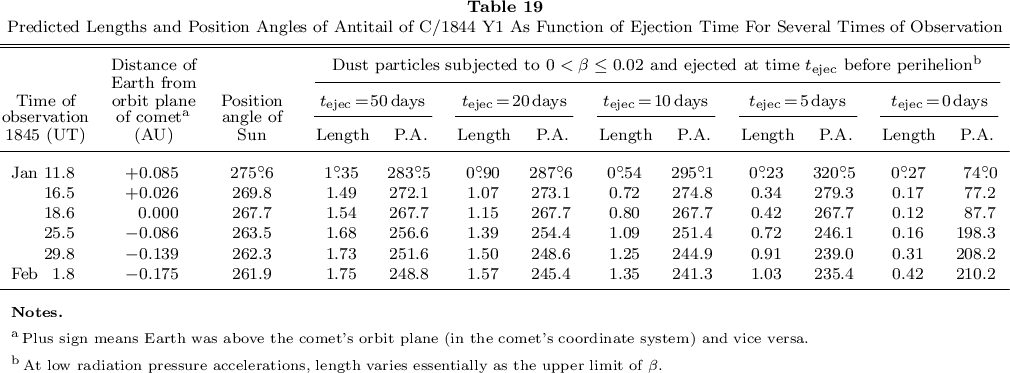}}} %  TABLE 19
\vspace{0.8cm}
\end{table*}

The meaning of Maclear's statements is clarified~\mbox{below} with help of
(i)~a plot of the dust ejecta's modeled distribution in the comet's orbital
plane, presented in Figure~9; and (ii)~the antitail's predicted length
and orientation as a function of the ejection time, shown in Table~19.
But first I offer the details of the other published account,~by Waterston
(1845), who reported from Bombay, India, that the comet displayed {\sl
``a singular luminous appendage, which, on the evening of [January] 16th}
(on about 16.55~UT) {\sl I observed for the first time to proceed from the
head of the comet {\sf towards\/} the Sun almost diametrically opposite to
the proper tail.  It consisted of a narrow band of faint light of about
the same breadth as the head.  The edges were well defined and parallel.
It could be traced for 3$^{\,\circ\!}$, and probably extended much
farther, as the increasing moonlight was very unfavorable to so faint
an object. \ldots \,[O]n the evening of [January] 25th [t]he two tails
\ldots made an evident angle, and the space was filled with a diffused,
irregular light, giving a triangular shape to the comet when seen by
the naked eye \ldots This evening} (the day not given, but before
January~31), {\sl the same appearance continues, but very faint, the
angle at the head of the comet being about 140$^{\,\circ\!}$.''}

Waterston also wrote a popular article about comet C/1844~Y1 and its
antitail in the {\it Bombay Courier\/}.  It was published on 1845
January~21 and copied by Pole (1845) in his paper on the comet.
While it does repeat some of the above lines, it is informative
enough to warrant a brief extract.  Waterston wrote:\ {\sl ``There is
a very remarkable appearance attending upon this Comet \ldots \,It
has got two tails!!  The second tail, which is nearly in an opposite
direction to the principal one, was first seen on \ldots 16~January
and has continued visible ever since, although the increasing
moonlight is very unfavorable.  It is extremely faint, but has been
recognized by several persons, and I have traced it distinctly for
about 3$^{\,\circ}$ towards the sun \ldots ''}

The attentive reader will note a major weakness of the narratives
by Maclear and Waterston in that they include no accurate data on the
orientation of the antitail and little information on its length.
Such data should narrow down the wide ranges of parametric values of
the model.  In fact, a few data of this kind --- specifically,
the position angles of the northern border of the antitail, which
Maclear referred to as the anterior luminous matter --- were published
in another paper by Maclear (1845c); the measurements were actually
made by his assistant at the Cape Observatory, W.\ Mann.  The issue
with these measurements is that they make no sense, a problem whose
ultimate solution brings me first to explaining the meaning of
Maclear's and Waterston's descriptions of the feature's observed
properties.

The systems of synchrones and syndynames in Figure~9 indicate that,
on the condition of a zero ejection velocity, the position of a
released dust particle at a given time depends on the time of
ejection, $t_{\rm ejec}$, and the magnitude of the solar radiation
pressure, $\beta$, it is subjected to.  The value of $\beta$ is a
function of the particle's physical and optical properties.  For
large particles (tens of microns and larger), $\beta$ varies
inversely as the product{\vspace{-0.04cm}} of their size and bulk
density.  For an assumed bulk density of 0.5~g~cm$^{-3}$, the
particle's effective diameter ${\cal D}_{\rm p}$ (in microns) equals
\begin{equation}
{\cal D}_{\rm p} = \frac{2.3}{\beta},
\end{equation}
where $\beta$ is in units of the solar gravitational acceleration.
The numbers in Table~19 apply to dust particles larger than
115~microns across.

\begin{table*}[t]
\vspace{0.17cm}
\hspace{0cm}
\centerline{
\scalebox{1}{
\includegraphics{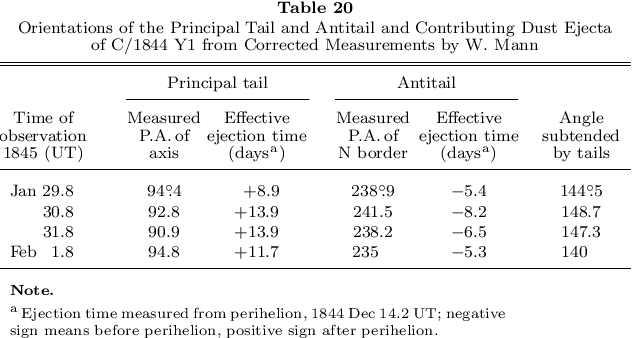}}} % TABLE 20
\vspace{0.66cm}
\end{table*}

The plot in Figure 9, depicting the field of dust ejecta in the
comet's orbital plane as it appeared at the time of Maclear's
first detection of the antitail, is remarkable in a number of
ways.  It often is stated that an antitail's sunward direction
is in its entirety an effect of projection and that in space the
antitail actually points away from the Sun.  Figure~9 shows that
this clearly is not the case here.  Particles ejected 50~days
before perihelion are at the time of observation obviously closer to
the Sun than the comet and the ones smaller than about 4~microns
across are actually located on the other side from the Sun.  Of
course these particles are too scattered in space to be seen.
It is only the large particles, having very low values of $\beta$
and coming mostly from near-perihelion ejections that Maclear
reported.  These in Figure~9 populate an area that is a little
farther from the Sun than the comet.  The dotted line connecting
the Earth with the comet's nucleus shows that, as seen on 11~January,
all large particles ejected earlier than about 2~days before
perihelion ended up in the antitail.

Since Maclear provided information on both the length and width of
the antitail on 11~January, one can learn additional information
on the ejecta from Table~19.  For example, if the dominant
contribution came from the dust ejected approximately 5~days
before perihelion, particles as small as about 20~microns in
diameter could have been located at the terminal point of the
feature.  If the prevailing ejecta were of much older age,
close to 80~days, the antitail should have contained only dust
grains whose diameter exceeded 120~microns.

No less interesting conclusion follows from Maclear's estimate of
the antitail's width, which can be interpreted in one of two ways.
Since 2$^\prime$ is equivalent to an arc of 1$^\circ\!$.5 at a
distance of 75$^\prime$, the estimated width could be generated by
the ejections continuing over a period of merely 0.3~day if the
dust was ejected 5 to 10~days before perihelion, but over some
11~days if between 20 and 50~days before perihelion.  Assuming
an exceptionally brief outburst, the same width should be
due to an ejection velocity normal to the orbit plane.  One then
obtains for the upper limit of this velocity a value between 6 and
16~m~s$^{-1}$, depending on the time of ejection.  In reality, the
outburst scenario is extremely unlikely and the interpretation in
terms of extended activity must be the correct one, implying a very
low ejection velocity, perhaps of about 1~m~s$^{-1}$ or so.  Its
true magnitude should more accurately be determined from the
antitail's width at the time of the Earth's transit across the
comet's orbital plane.  Unfortunately, the feature was then
invisible on account of the strong moonlight.

The adopted explanation of the antitail's width as an effect of
continuous production of dust is also consistent with Waterston's
account of the comet's appearance on 25~January, when he noted
the diffuse light filling the space between the antitail and the
principal tail.  This feature confirms that the emission of dust
was continuous and that it was missed by Maclear only because of
the ejecta's extremely low surface brightness.

Related to this issue is the comment that Maclear made with respect
to his observation of the antitail's borders on 27~January in the
comet sweeper and the large telescope.  He noted that the northern
border was sharply defined, whereas the southern border was diffused.
From Table~19 it is obvious that the northern border was made up of
the earliest ejecta that the observer was able to detect, whereas
the southern border consisted of the most recent ejecta that he
could recognize, given the surface-brightness thresholds of his
instruments.  It should be emphasized that before the Earth crossed
the comet's orbital plane, the roles of the two borders had been
interchanged:\ the southern border had been then the one populated
by the earliest ejecta.

I now come back to the controversial position angle observations
by Mann (Maclear 1845c), the only ones of the kind that I was able
to find.  The published data refer to the axis of the principal
tail and to the northern (sharper) border of the antitail on each
of the four nights between 29~January and 1~February.  While it
is remarkable that the fearure could still be measured at all two
weeks after the orbital plane's transit, Mann's nominal position
angles of the antitail's border on the four nights, equaling
301$^\circ\!$.1, 298$^\circ\!$.5, 301$^\circ\!$.8, and 305$^\circ$,
respectively, cannot be correct, because the northern border was
in the third, not fourth, quarter.  If Mann measured the position
angle from the parallel, it could be that the term ``northern''
misled him to erroneously add --- rather than subtract --- the
angle that the border subtended with the direction to the west.
If so, the correct position angles of the antitail's northern
border, PA$_{\rm corr}$, are related to its published values,
PA$_{\rm publ}$, by \mbox{PA$_{\rm corr} = 540^\circ -
{\rm PA}_{\rm publ}$}.  That this explanation of the meaningless
numbers is plausible appears to be documented by comparing the
angle of 140$^\circ$ between the two tails, reported at
approximately the same time by Waterston, with the same angle,
computed from Mann's measurements to equal 144$^\circ\!$.5
after correction but 153$^\circ\!$.3 before it.

Table 20 lists Mann's measured position angle of the principal
tail's axis, the corrected position angle of the antitail's
northern border, the angle they subtend, and the effective
times of ejection corresponding to the respective position
angles.  These times were derived from the computed position
angles of a system of synchrones.  The telescopic measurements,
which obviously referred to the orientations of the axis and
border near the comet's nucleus, were compared to the model
values at very low accelerations $\beta$.  The results, based
on the observations made between 29~January and 1~February,
suggest that the earliest ejecta in the {\it observed\/} antitail were
large grains (mostly exceeding 100~microns in diameter), released,
on the average, about 5~days before perihelion, whereas the
major contribution to the principal tail was the microscopic
(micron-sized) dust ejected at times centered approximately
on 12~days after perihelion.  Accordingly, much of the material
in the two tails and in space between them derived from dust
ejections that extended over a little longer than two weeks
asymmetrically centered on perihelion.

Finally, the antitail observations provide evidence of the
comet's fairly high rate of dust production before perihelion.
One may have questioned the level of this activity, given that
the comet was not discovered until after perihelion.  However,
the reason for the failure was the extremely unfavorable
preperihelion observing conditions, with the comet staying
at solar elongations not exceeding 35$^\circ$ between 3~AU
from the Sun in early July 1844 and perihelion.

\section{Pair of C/1844 Y1 and C/2019 Y4 as Extreme Case Among Known
 Long-Period Comet Groups}
The pair of C/1844 Y1 and C/2019 Y4 extends~the~very limited number of
groups of long-period comets in nearly identical orbits, undoubtedly
products of relatively recent nucleus fragmentation.  The largest and
best known among these groups is a foursome headed by the principal
fragment C/1988~A1 (Liller) and comprising the companion fragments
C/1996~Q1 (Tabur), C/2015~F3~(SWAN), and C/2019~Y1 (ATLAS).  It was
sheer coincidence that the group's most recent member happened to be
discovered by the \mbox{ATLAS} project less than two weeks prior to
C/2019~Y4.

The Liller group is the only one known to have more than two members.
Before the discovery of C/2019~Y4, the best defined pair consisted
of C/1988~F1 (Levy), the principal fragment, and C/1988~J1
(Shoemaker-Holt), the companion fragment.  Of the two additional,
%
% The intrinsically brightest object always arrived at perihelion first,
% the companion's record time lag of 175.5~yr being held by C/2019~Y4.
%
probable pairs that should be mentioned, one includes C/1915~R1
(Mellish) and C/2016~R3 (Borisov) and is unusual in more than one
respect. The orbits differ from each other a little more than in the
other instances, but this could be because of a poorly determined
orbit of comet Mellish (Williams 2016).  Rather peculiar, but
entirely coincidental is that the two objects arrived at the same
time of the year to within three days under equally unfavorable
conditions.  Both were astrometrically observed over very limited
periods of time before perihelion, Mellish over just 4~days, Borisov
some two weeks.  Neither was picked up after perihelion, although
under ordinary circumstances they should have been, in the southern
hemisphere (Einersson 1915, Aitken 1915, Green 2016).  It is significant
that attempts to link the motions of the two comets failed (Williams
2016), so these were not returns of the same object.  It is possible
that neither Mellish nor Borisov survived perihelion and that ---
unlike in the above three examples --- {\it both comets were companion
fragments to an unknown principal fragment\/} that had passed perihelion
unobserved many decades before 1915.  An approximate original orbital
period for comet Borisov equals 2330\,$\pm$\,880~yr.\footnote{See {\tt
https://minorplanetcenter.net/db\_search.}}

The other probable pair is linked to the story of the so-called Lick
Object, a very unusual phenomenon observed near the setting Sun by a
party at the residence of the Lick Observatory Director W.\ W.\ Campbell
on 7~August 1921.  The event, described in the literature (Campbell
1921a, 1921b) and corroborated by a few independent reports, was
believed to be most probably an unknown comet.  A recent detailed
investigation by Sekanina \& Kracht (2016) led to a conclusion
that the Lick Object apparently was a fragment of a comet
that it had shared with the sunskirter C/1847~C1 (Hind).

\subsection{Fragments' Returns to Perihelion and\\[-0.01cm] Orbital
 Periods\\[-0.1cm]}
The data on the groups of long-period comets, presented in Table~21
to be discussed in Section~7.2, display some interesting features.
The term {\it long period\/} is defined by the range of original
orbital periods of the four groups, whose principal fragments are
known.  According to the catalogue by Marsden \& Williams (2008), the
periods varied from $\sim$2900~yr for comet Liller to $\sim$14,000~yr
for comet Levy.  The Great Comet of 1844 and comet Hind had
intermediate periods of 7600~yr and 8300~yr, respectively.  The
typical original orbital period is thus on the order of several
thousand years, with a maximum-to-minimum ratio~of~approximately
5:1.\vspace{-0.01cm}

The appearance of the principal and companion fragments, the order
of return to perihelion, and the range of time lags between the
principal and companion fragments among these long-period comets
vary so strikingly that there must be major reasons for the differences.
The principal fragments are intrinsically much brighter than the
companions and their light curves are, to the extent known, more or
less symmetrical with respect to perihelion.  By comparison, the
companion fragments usually exhibit a steep drop in brightness
after perihelion, if they survive that long.  The process of
disintegration for at least two --- C/1996~Q1 and C/2019~Y4 ---
began early and their demise may have been essentially completed
by the time of perihelion or shortly afterwards.\vspace{-0.01cm}

The principal fragment always returned to perihelion first, a
circumstance that has a diagnostic value as proposed below.~This
rule is known to also apply~to~the~non\-tidally split comets with
persistent companions~in~the short-period orbits, e.g., 3D/Biela,
73P/Schwassmann-{\linebreak} Wachmann, 79P/du Toit-Hartley,
141P/Machholz, etc. (see Marsden \& Williams 2008).\vspace{-0.01cm}

The range of time lags of the companion fragments trailing behind
the principal fragments at perihelion is astonishing, extending from
as little as 0.21~yr for the pair of C/1988~F1 and C/1988~J1 to 175.5~yr
for the pair of C/1844~Y1 and C/2019~Y4, nearly three orders of magnitude.
Persistent companions of the split comets in short-period orbits, such
as 3D, 73P, or 141P, are known to pass perihelion a few hours to about
one day after the principal fragment, which is on the order of hundredths
of one percent of the orbital period.  On the other hand, the lag for
C/2019~Y4 equaled fully 2.3~percent of the orbital period of C/1844~Y1,
if not more.

When a comet in a long-period orbit splits into two fragments
that survive a full revolution about the Sun, the companion
fragment arrives at its perihelion at a time different from the
perihelion time of the principal fragment.  This differential
effect could be triggered by (i)~the separation velocity, (ii)~the
sublimation-driven nongravitational acceleration, and/or (iii)~the
indirect planetary perturbations (even in the absence of either
fragment's close encounter with a planet).\vspace{-0.01cm}

\begin{table*}[t]
\vspace{0.17cm}
\hspace{-0.17cm}
\centerline{
\scalebox{0.97}{
\includegraphics{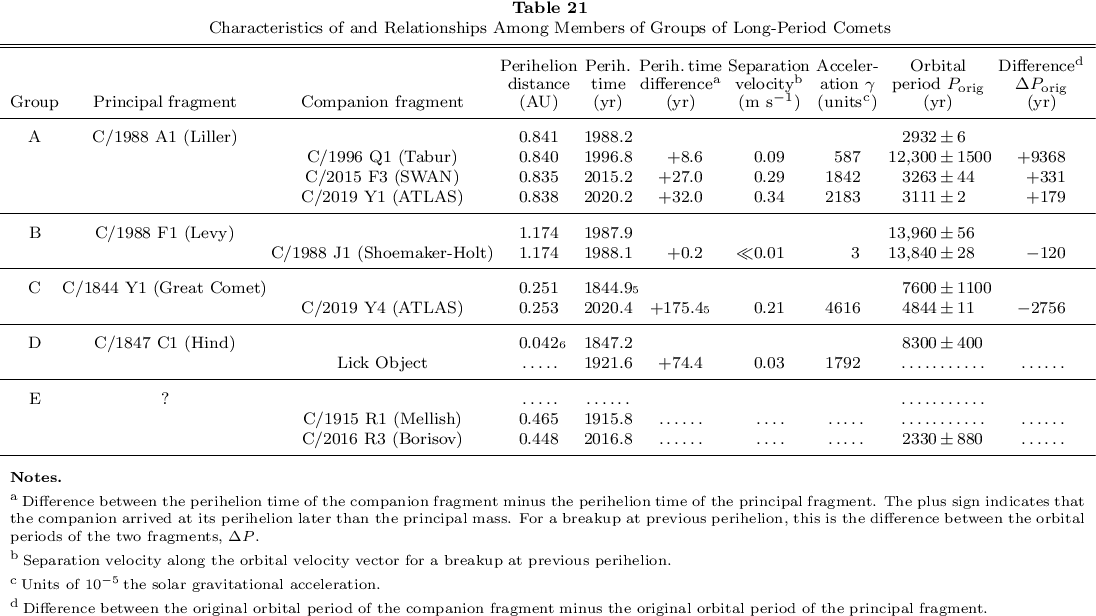}}} %  TABLE 21
\vspace{0.6cm}
\end{table*}

I now investigate the magnitude of the differential effect in the
orbital period for each of the three categories of provenance.  I
assume that the companion separates from the parent in the general
proximity or perihelion.  Differentiating the expression for the
orbital velocity $V_{\rm orb}$ as a function of heliocentric
distance $r$ at a fixed location of the fragmentation event, given
by $r_{\rm frg}$, the relation between the separation velocity,
$V_{\rm sep}(r_{\rm frg})$, measured along the orbital-velocity
vector, and the corresponding relative change in the orbital period,
$\Delta P/P$, is
\begin{equation}
V_{\rm sep}(r_{\rm frg}) = \frac{k_0^2(1 \!-\! e)}{3q V_{\rm orb}
 (r_{\rm frg})} \frac{\Delta P}{P},
\end{equation}
where $q$ is the perihelion distance of the parent comet, $e$ its
eccentricity, and $k_0$ the Gaussian gravitational constant.  In
a pseudo-parabolic approximation,
\begin{equation}
V_{\rm sep}(r_{\rm frg}) = \kappa \sqrt{{\rule{0ex}{1.4ex}}r_{\rm frg}} \,
 P^{-\frac{5}{3}} \Delta P,
\end{equation}
where the constant $\kappa$ equals
\begin{equation}
\kappa = {\textstyle \frac{1}{3}} \, 2^{\frac{1}{6}}
 \pi^{\frac{2}{3}} k_0^{\frac{1}{3}}.
\end{equation}
When $r_{\rm frg}$ is given in AU, $P$ and{\vspace{-0.04cm}} $\Delta P$
in years, and $V_{\rm sep}$ in m~s$^{-1}$, the constant is numerically
\mbox{$\kappa = 7020$}.

The differential effect of a nongravitational acceleration consists
in the fact that the fragments orbit the Sun in gravity fields of
slightly unequal magnitudes.  The effect on the orbital period,
$P$, follows from Kepler's Third Law
\begin{equation}
P = \frac{2 \pi}{k_0} \, a^{\frac{3}{2}},
\end{equation}
where $a$ is the semimajor axis.  Differentiating this equation with
respect to $k_0$, which is no longer a constant, one finds
\begin{equation}
\Delta P = -\frac{2 \pi}{k_0^2} \Delta k_0 \, a^{\frac{3}{2}},
\end{equation}
so that
\begin{eqnarray}
\frac{\Delta P}{P} & = & -\frac{\Delta k_0}{k_0} = -\frac{\Delta k_0^2}{2k_0^2}
	= {\textstyle \frac{1}{2}} \, 10^{-5} \gamma.\\[-0.1cm]
	& & \nonumber
\end{eqnarray}
Here $\gamma$ is the parameter of the nongravitational model applied
in Section~3 to investigate the fragmentation of C/2019~Y4; see also
Equations~(10) and following.{\vspace{-0.03cm}}  The units of $\gamma$
have been $10^{-5}$ the solar gravitational acceleration.

Even though the planets gravitationally affect the motions of comets in
all six elements, the perturbations of the reciprocal semimajor axis,
$1/a$, have always been of primary interest because they present changes
in the orbital energy and are relevant to the issues of comets' perihelion
returns or escape from the Solar System.  The indirect perturbations of
$1/a$ of long-period comets were investigated extensively by Everhart \&
Raghavan (1970).  Everhart (1968) also undertook a Monte Carlo study
of perturbations by a single planet, concluding that their distribution
was not Gaussian.

As two long-period fragments are on their way to aphelion, the indirect
planetary perturbations essentially cease at heliocentric distances smaller
than 50~AU.  By that time the separation distance between the two objects
has not increased enough to imply a major differential perturbation
effect.  However, even in the absence of a significant contribution from a
separation velocity and/or a nongravitational acceleration, the aphelion
distances of the fragments are not identical.  It is near aphelion, where
their separation distance could increase dramatically, so that the original
semimajor axes of the fragments determined from their subsequent
near-perihelion arcs could become clearly uneven because of the unequal
integrated effects of the indirect planetary perturbations at
heliocentric distances smaller than 50~AU.

Changes in the original orbital energy (i.e., its value near
aphelion, referred to the barycenter of the Solar System),
$\Delta (1/a)_{\rm orig}$, are convertible into changes in the
original orbital period, $P_{\rm orig}$.  Since Kepler's
Third Law can obviously be written as
\begin{equation}
P_{\rm orig} = \frac{2 \pi}{k_{\rm b}} (1/a)_{\rm orig}^{-\frac{3}{2}},
\end{equation}
with $k_{\rm b}$ being the ``barycentric'' Gaussian constant, the
differentiation with respect to $(1/a)_{\rm orig}$ gives
\begin{equation}
\Delta P_{\rm orig} = -\frac{3 \pi}{k_{\rm b}} \, a_{\rm
 orig}^{\frac{5}{2}} \, \Delta (1/a)_{\rm orig}
\end{equation}
and, dividing Equation (29) by Equation (28),
\begin{equation}
\Delta P_{\rm orig} = - {\textstyle \frac{3}{2}} P_{\rm orig} a_{\rm
 orig} \, \Delta (1/a)_{\rm orig}
 = - {\textstyle \frac{3}{2}} P_{\rm orig}^{\frac{5}{3}}
 \Delta (1/a)_{\rm orig},
\end{equation}
where the expression on the right applies only when $P_{\rm orig}$
--- and $\Delta P_{\rm orig}$ --- are reckoned in years.  Instead
of differentiation one can work directly with the differences.

\subsection{Numerical Results for the Individual Groups}

I now assess the three proposed sources for $\Delta P$ and
$\Delta P_{\rm orig}$.  For the sake of simplifying the matters,
I assume that the parent comet fragmented at perihelion.  The
results in Table~21 offer surprisingly robust conclusions about
effects of the separation velocity, the nongravitational acceleration,
and the indirect planetary perturbations on the returns of the
fragments to perihelion.

First of all, it is not accidental that the orbital periods of comets
in the groups or pairs are not longer than thousands of years.  If
they were many tens of thousands of years or longer, the fragments
would have been returning over periods in the past well beyond our
records of cometary orbits.  In fact, Table~21 illustrates that we
already begin to have problems of this kind (such as the missing
principal fragment of Group~E).  In this context, it is possible,
for example, that the disintegrating comet C/1999~S4 (\mbox{LINEAR})
was a companion to a principal fragment whose orbital period may
have been, say, 100,000~yr or so and had returned to perihelion a
number of centuries earlier, as I already suggested (\mbox{Sekanina}
2000).  C/1999~S4 masqueraded as a potential Oort Cloud comet (see
footnote~1 on page~19), but was strongly deficient in carbon monoxide
ice (Weaver et al.\ 2001), which helped reveal its true origin.

Separation velocities can hardly be the source of the investigated
effect on companion fragments that are all {\it trailing\/} the principal
fragments.  On the contrary, one expects that effects of the separation
velocity should lead to statistically equal numbers of companion
fragments preceding and following the principal fragments.  Even though
the dataset in Table~21 is very small, the absence of leading companions
is striking.  Also suspect are the minuscule magnitudes of the
derived separation velocities; one would have to assume that {\it
all\/} fragmentation events occurred far from the Sun, thereby
dramatically increasing the effect of $r_{\rm frg}$ in Equation~(23).

Effects of the nongravitational acceleration are consistent with the
dominance of trailing companion fragments, but the obvious problem is
the gigantic magnitude of the acceleration required (with the exception
of C/1988~J1).  Objects with such enormous nongravitational efffects
could not appear as observable objects and survive the entire
revolution about the Sun.  There is also no correlation between the
likelihood of survival and the magnitude of $\Delta P$ in Table~21.
For example, if the nongravitational acceleration governed this
relationship, one would expect C/2019~Y1 much more likely to have
begun disintegrating before perihelion than C/1996~Q1, contrary to
the observed behavior of the two comets.  Yet, it should be admitted
that the nongravitational acceleration {\it does to a degree
contribute\/} to the distribution of perihelion returns among the
companion fragments, tilting $\Delta P$ to positive values in
Table~21.

This leaves the indirect planetary perturbations as the major source
of the effects that influence the semichaotic timing of the companion
fragments' perihelion returns.  The three well-established
groups/pairs in Table~21 suggest a strong dependence on the
perihelion distance.~One could also note that this empirical
finding fits in with~a pair of the Kreutz sungrazers, even
though these comets have distinctly shorter orbital periods.
While the perihelion returns of the companion fragments of the
Liller group lagged the principal fragment, whose perihelion
distance amounted to 0.84~AU, by 0.3~percent to 1.1~percent of the
principal fragment's orbital period, C/2019~Y4 trailed C/1844~Y1,
whose perihelion distance was 0.3 the value of C/1988~A1, by
at least 2.3~percent of its orbital period.  By comparison, the
Kreutz sungrazer C/1965~S1 (Ikeya-Seki), a less massive sibling
of C/1882~R1 (Great September Comet), trailed the latter at
perihelion by $>$11~percent of its orbital period.

In any event, the pair of C/1844~Y1 and C/2019~Y4 is an extreme
case among the known groups/pairs of long-period comets.  If the
tabulated original orbital period of 7600~yr for C/1844~Y1
(Marsden et al.\ 1978) is replaced by T.\ Kobayashi's value
of 4000~yr (Green 2020a), the perihelion return of C/2019~Y4
would have lagged that of the principal fragment by as much
as 4.4~percent of its orbital period, but still by significantly
less than in the case of the Kreutz pair.  And even though the
value of $\Delta P_{\rm orig}$ would change from $-$2756~yr to
a positive value of +844~yr, the unpublished 1$\sigma$ uncertainty
in the orbital period of C/1844~Y1 derived by Kobayashi is estimated
at hardly much less than $\pm$40~percent or about $\pm$1500~yr
(given that the orbital solution had a high mean residual of
$\pm$5$^{\prime\prime}\!$.2), making $\Delta P_{\rm orig}$ for C/2019~Y4
essentially indeterminate.  Further contributing to the uncertainty
is the error of $P_{\rm orig}$ of C/2019~Y4 itself.  Because
original orbital periods derived from nongravitational solutions
are known to be notoriously unreliable (Marsden et al.\ 1973),
the tabulated $P_{\rm orig}$ for C/2019~Y4 was taken from one of
the last gravitational-orbit determinations, before introducing
the nongravitational terms into the equations of motion became
inevitable.

\section{Conclusions}
The present investigation addresses five issues associated
with the genetically related pair of long-period comets, C/1844~Y1
and C/2019~Y4.  The first, the Great Comet of 1844 (or 1844/1845),
was a naked-eye object that burst into sight for southern-hemisphere
observers a few days after perihelion in the second half of December
1844 and later was being monitored telescopically until late March
1845.  The second comet was discovered about five months before
perihelion and kept observers guessing about its next move.  It
brightened rapidly in late January 2020 only to eventually
disintegrate before reaching perihelion at the end of May.

I first examine the brightness of C/2019 Y4:\ (i)~the light curve of
its nuclear condensation from the time of a prediscovery observation
(20~December 2019) to late February~2020; and (ii)~its integrated
brightness (both visual and CCD) between late February and late May.
Intrinsically, the nuclear condensation was fading until about 22~January,
when this trend was reversed and the comet began to brighten rapidly,
at a rate of 0.11 mag per day.  The total apparent brightness reached
a broad peak in the last days of March and began to subside at a very
slow rate.  The light curve then essentially stabilized according to
some observers or continued to decline after another minor peak
according to other observers.  In May the comet became a difficult
object, its solar elongation dropping below 30$^\circ$ on the 17th.

A major part of this paper deals with the fragmentation of C/2019~Y4.
The separation vectors involving the four fragments, A, B, C, and D,
introduced by the MPC, are tested as a function of time, the fragment
identities are examined and where needed corrected.  Application of the
standard model confirms that Fragment~B was the principal (and
presumably the most massive) fragment.  The iterations of the time of
separation of B and A have failed to converge, but the distribution
of residuals has gradually been improving with the separation moving
to earlier times.  From this standpoint, the breakup in March is
clearly inadmissible, but equating the time of breakup with the
onset time of the nuclear-condensation's rapid brightening around
22~January offers an acceptable solution (BA$_1$; Table~3), besides
making sense in terms of physics.  Fragment~A was short lived and
subject to a nongravitational acceleration of about 20~units of
10$^{-5}$ the solar gravitational acceleration relative to Fragment~B.

The other two MPC-recognized fragments, C and D, are likely to have
split off from Fragment~A in mid-March (solutions AC$_2$ in Table~5
and AD$_3$ in Table~13, respectively).  Their motions were affected
by considerably higher nongravitational accelerations, in excess of
100~units of 10$^{-5}$ the solar gravitational acceleration relative
to Fragment~A.  They also belonged to the category of short-lived
fragments and their lifespans were, as expected, much shorter than
the lifespan of Fragment~A.

The separation data of A and another fragment imaged with the
Haleakala-Faulkes Telescope North (code F65) on 17 and 18~April
could not be assigned to any of the known pairs of fragments,
but could be linked with the separation from A of the fragment
measured at Osservatorio del Celado (code K51) on 9~April that
likewise could not be identified with any of the other fragments.
These observations are proposed to refer to a new, fifth Fragment E.
Based on a correlation between sizes and accelerations for the split
comets, it is estimated that the pre-split nucleus of C/2019 Y4
was initially 500~meters across, Fragment~A 100~meters, and
Fragments~C, D, an E at most 20~meters.

The relative trajectories of the five fragments projected onto
the plane of the sky are briefly studied.  The most intriguing
is the looped trajectory of Fragment A relative to B; the
angular distance between the two fragments reached a maximum
of 7$^{\prime\prime}\!$.5 on 19~March, after which time they
were gradually approaching each other until 19~April, when
they were merely 3$^{\prime\prime}\!$.4 apart.  From that
time on, the distance began to increase again, reaching
8$^{\prime\prime}\!$.7 on 1~May and 17$^{\prime\prime}\!$.4
a week later.  Until early April, the only visible fragment
was A, which was very active ever since the time of breakup in
late January.

The third item discussed is the history of orbit determination
of C/1844~Y1.  The first major~result was a hyperbolic orbit
by Bond (1850), which was shown to be erroneous by Fayet (1910).
The corrected orbit was an ellipse with an orbital period of
nearly 30,000~yr, fitting the normal places only slightly
better than a parabola.  My orbit from 1974, based on 41~most
consistent astrometric observations (a mean residual of
$\pm$1$^{\prime\prime}\!$.8) and later incorporated into
Marsden et al.'s (1978) paper, yielded an original orbital period
of 7600\,$\pm$\,1100~yr.  By contrast, T.\ Kobayashi's recently
derived orbit (Green 2020a), based on a different set of
41~observations (a mean residual of $\pm$5$^{\prime\prime}\!$.2),
gave 4000~yr (1$\sigma$ uncertainty not provided).  I conclude
that the orbital period of C/1844~Y1 equals several thousand
years and cannot be determined more accurately because
of low accuracy of the astrometric observations and short orbital
arc available.

The fourth item addressed is the antitail of C/1844~Y1, displayed
in January 1845 thanks to favorable projection conditions.
Unfortunately, the moonlight was interfering on and around 18~January,
the day of the Earth 's crossing the comet's orbit.  The full Moon
was five days later.  Useful observations of the antitail were
reported from two locations, the Royal Observatory, Cape of Good
Hope, and Bombay, India, both on days before the transit and in
late January, long after the transit.  Interestingly, one of the
accounts noted that space between the antitail and main tail was
filled with a diffused, irregular light, as predicted by the theory.

Limited analysis suggests that the antitail contained dust
grains several tens of microns in diameter near its sunward end
and much larger ones closer to the nucleus, ejected from it
before perihelion.  Position angle measurements of the
antitail's northern border and main tail's axis suggest
that the optically dominant dust ejecta left the nucleus
between 5~days before perihelion and 12~days after perihelion.

The last item is examination of the genetically related pair of C/1844~Y1
and C/2019~Y4 as an extreme case among other such groups/pairs of
long-period comets, including comparison with the Liller
group of four kindred objects:\ C/1988~A1, C/1996~Q1, C/2015~F3,
and C/2019~Y1.  Subject to future confirmation by additional
examples, each group/pair appears to be a product of a fragmentation
event (or possibly events in the case of groups) about the time of
previous perihelion (typically several thousand years ago).  One
(principal) fragment is intrinsically much brighter than the other
(companion) fragments, some of which were observed failing to survive
through perihelion, others fading precipitously after perihelion.
The principal fragment always returns first, the companion fragment(s)
following a fraction of a year to nearly two hundred years later.
Of three potential trigger mechanisms proposed to explain the time
lags of the trailing companions, the dominant effect is believed
to be provided by the indirect planetary perturbations, with a
minor but instrumental contribution from the sublimation-driven
nongravitational effect.  The separation velocity at breakup is
suggested to be inconsequential because it could not explain the
absolute prevalence of trailing companion fragments.
%
% \pagebreak
%

%
\begin{center}
{\footnotesize REFERENCES}
\end{center}
\vspace{-0.15cm}
\begin{description}
{\footnotesize
\item[\hspace{-0.3cm}]
Aitken, R.\ G.\ 1915, PASP, 27, 244
\\[-0.57cm]
\item[\hspace{-0.3cm}]
Bond, G.\ P.\ 1850, AJ, 1, 97 % 138 obs.
\\[-0.57cm]
\item[\hspace{-0.3cm}]
Campbell, W.\ W.\ 1921a, Nature, 107, 759
\\[-0.57cm]
\item[\hspace{-0.3cm}]
Campbell, W.\ W.\ 1921b, PASP, 33, 258
\\[-0.57cm]
%
% \item[\hspace{-0.3cm}]
% Curtis, H.\ D.\ 1911, PASP, 23, 122 % Fayet found "serious error" by Bond.
% \\[-0.57cm]
%
\item[\hspace{-0.3cm}]
Einarsson, S.\ 1915, PASP, 27, 243
\\[-0.57cm]
\item[\hspace{-0.3cm}]
Everhart, E.\ 1968, AJ, 73, 1039
\\[-0.57cm]
\item[\hspace{-0.3cm}]
Everhart, E., \& Raghavan, N.\ 1970, AJ, 75, 258
\\[-0.57cm]
\item[\hspace{-0.3cm}]
Fayet, G.\ 1910, Ann.\ Obs.\ Paris, 26A, 1 % Re 1844 III, p. 59, 111-118; sign
\\[-0.57cm]
\item[\hspace{-0.3cm}]
Green, D.\ W.\ E.\ 2016, CBET 4321
\\[-0.57cm]
\item[\hspace{-0.3cm}]
Green, D.\ W.\ E.\ 2020a, CBET 4712
% discovery of C/2019 Y4
\\[-0.57cm]
\item[\hspace{-0.3cm}]
Green, D.\ W.\ E.\ 2020b, CBET 4708
% discovery of C/2019 Y1
\\[-0.57cm]
\item[\hspace{-0.3cm}]
Green, D.\ W.\ E.\ 2020c, CBET 4744
\\[-0.57cm]
\item[\hspace{-0.3cm}]
Green, D.\ W.\ E.\ 2020d, CBET 4751 %  (see also 4763)
\\[-0.57cm]
\item[\hspace{-0.3cm}]
Hui, M.-T., \& Ye, Q.-Z.\ 2020, AJ, 160, 91 (10pp)
% comet has disintegrated since mid-March; A1 = 22.5 x 10^-8
\\[-0.57cm]
%
% \item[\hspace{-0.3cm}]
% James, N.\ 2020, J.\ Brit.\ Astron.\ Assoc., 130, 133
% astrometric residuals show gradual divergence after Mar 25
% \\[-0.57cm]
%
\item[\hspace{-0.3cm}]
Knight, M., \& Battams, K.\ 2020, Astron.\ Telegram, 13813
\\[-0.57cm]
%
% \item[\hspace{-0.3cm}]
% Lin, Z.-Y., Hsia, C.-H., \& Ip, W.-H.\ 2020, Astron.\ Telegram, 13629
% at least 2 fragments on Apr 12
% \\[-0.57cm]
%
\item[\hspace{-0.3cm}]
Maclear, T.\ 1845a, Mon.\ Not.\ Roy.\ Astron.\ Soc., 6, 213
\\[-0.57cm]
\item[\hspace{-0.3cm}]
Maclear, T.\ 1845b, Mon.\ Not.\ Roy.\ Astron.\ Soc., 6, 234
\\[-0.57cm]
\item[\hspace{-0.3cm}]
Maclear, T.\ 1845c, Mon.\ Not.\ Roy.\ Astron.\ Soc., 6, 252
\\[-0.57cm]
%
% \item[\hspace{-0.3cm}]
% Minor Planet Center Staff 2020a, MPEC 2020-A72
% \\[-0.57cm]
%
% \item[\hspace{-0.3cm}]
% Minor Planet Center Staff 2020b, MPEC 2020-J16
% \\[-0.57cm]
%
\item[\hspace{-0.3cm}]
Marcus, J.\ N.\ 2007, Internat.\ Comet Quart., 29, 39
\\[-0.57cm]
\item[\hspace{-0.3cm}]
Marsden, B.\ G.\ 1972, Catalogue of Cometary Orbits, 1st ed.,{\linebreak}
 {\hspace*{-0.6cm}}Smithsonian Astrophysical Observatory, Cambridge,
 MA, 70pp
\\[-0.35cm]
\pagebreak
\item[\hspace{-0.3cm}]
Marsden, B.\ G.\ 1975, Catalogue of Cometary Orbits, 2nd ed.,{\linebreak}
 {\hspace*{-0.6cm}}Smithsonian Astrophysical Observatory, Cambridge, MA,
 83pp
\\[0.9cm]
\item[\hspace{-0.3cm}]
Marsden, B.\ G., \& Williams, G.\ V.\ 2008, Catalogue of Cometary{\linebreak}
 {\hspace*{-0.6cm}}Orbits, 17th ed., IAU Minor Planet Center/Central Bureau
 for{\linebreak}
 {\hspace*{-0.6cm}}Astronomical Telegrams, Cambridge, MA, 195pp
\\[-0.57cm]
\item[\hspace{-0.3cm}]
Marsden, B.\ G., Sekanina, Z., \& Everhart, E.\ 1978, AJ, 83, 64
\\[-0.57cm]
\item[\hspace{-0.3cm}]
Marsden, B.\ G., Sekanina, Z., \& Yeomans, D.\ K.\ 1973, AJ,~78,~211
\\[-0.57cm]
\item[\hspace{-0.3cm}]
Minor Planet Center Staff 2020, MPEC 2020-K131 \& 2020-J16
\\[-0.57cm]
%
% \item[\hspace{-0.3cm}]
% Nakano, S.\ 2020, CBET 4744
% NG effects beginning around Mar 31
% \\[-0.57cm]
%
\item[\hspace{-0.3cm}]
Pole, W.\ 1845, J.\ Bombay Branch Roy.\ Asiat.\ Soc., 2, 201
\\[-0.57cm]
\item[\hspace{-0.3cm}]
Sekanina, Z.\ 1973, IAU Circ.\ 2580
\\[-0.57 cm]
\item[\hspace{-0.3cm}]
Sekanina, Z.\ 1974a, Icarus, 23, 502
\\[-0.57cm]
\item[\hspace{-0.3cm}]
Sekanina, Z.\ 1974b, Sky Telesc., 47, 374
\\[-0.57cm]
%
% \item[\hspace{-0.3cm}]
% Sekanina, Z.\ 1975, Icarus, 25, 218
% \\[-0.57cm]
%\
\item[\hspace{-0.3cm}]
Sekanina, Z.\ 1976, in Interplanetary Dust and Zodiacal Light,{\linebreak}
 {\hspace*{-0.6cm}}IAU Coll.\ 31, ed.\ H.\ Els\"{a}sser \& H.\ Fechtig
 (Berlin:\ Springer), 339
\\[-0.57cm]
\item[\hspace{-0.3cm}]
Sekanina, Z.\ 1977, Icarus, 30, 574
\\[-0.57cm]
\item[\hspace{-0.3cm}]
Sekanina, Z.\ 1978, Icarus, 33, 173
\\[-0.57cm]
\item[\hspace{-0.3cm}]
Sekanina, Z.\ 1982, in Comets, ed.\ L.\ L.\ Wilkening (Tucson:\ Univ.{\linebreak}
 {\hspace*{-0.6cm}}Arizona), 251
\\[-0.57cm]
\item[\hspace{-0.3cm}]
Sekanina, Z. 2000, IAU Circ.\ 7471
\\[-0.57cm]
\item[\hspace{-0.3cm}]
Sekanina, Z.\ 2010, Int.\ Comet Quart., 32, 45
\\[-0.57cm]
\item[\hspace{-0.3cm}]
Sekanina, Z., \& Kracht, R.\ 2016, ApJ, 823, 2 (26pp)
\\[-0.57cm]
\item[\hspace{-0.3cm}]
Sekanina, Z., \& Miller, F.\ D.\ 1976, Icarus, 27, 135
\\[-0.57cm]
%
% \item[\hspace{-0.3cm}]
% Stefanik, R.\ P.\ 1966, M\'em.\ Soc.\ Roy.\ Sci.\ Li\`ege
%  (S\'er.\ 5), 12, 29
% \\[-0.57cm]
%
\item[\hspace{-0.3cm}]
Waterston, J.\ J.\ 1845, Mon.\ Not.\ Roy.\ Astron.\ Soc., 6, 207
\\[-0.57cm]
\item[\hspace{-0.3cm}]
Weaver, H.\ A., Sekanina, Z., Toth, I., et al.\ 2001, Science, 292, 1329
\\[-0.57cm]
\item[\hspace{-0.3cm}]
Whipple, F.\ L.\ 1950, ApJ, 111, 375
\\[-0.57cm]
\item[\hspace{-0.3cm}]
Williams, G.\ V.\ 2016, MPEC 2016-S03
\\[-0.57cm]
\item[\hspace{-0.3cm}]
Ye, Q., \& Zhang, Q.\ 2020, Astron. Telegram, 13620
% elongated pseudo-nucleus on Apr 5
\\[-0.64cm]
\item[\hspace{-0.3cm}]
Ye, Q., Jewitt, D., Hui, M.-T., et al.\ 2021, AJ, 162, 70 (13pp)}
% first major disruption -60 days
%
\vspace{0.76cm}
%
% C/1915 R1  118.8  73.9  53.5  0.443  1.0      Mellish  PASP 27, 243, 244
% C/2016 R3  118.0  78.7  53.0  0.447  0.9955   Borisov  CBET 4321; 2330+/-880
% C/1915 R1  116.2  77.5  52.3  0.465  1.0      new orbit MPEC 2016-S03
%
\end{description}
\end{document}